\documentclass[12pt]{article}
\usepackage[left=2.5cm, right=2.5cm, top=3cm, bottom=3cm]{geometry}
\usepackage{appendix}
\usepackage[utf8]{inputenc}
\usepackage{setspace}
\usepackage[]{amsmath,bm}
\usepackage{amssymb}
 \newcommand{\sym}[1]{$^{#1}$}
\usepackage{natbib}
\usepackage{appendix}
\usepackage[table,usenames,dvipsnames]{xcolor}
\definecolor{burgundy}{rgb}{0.5, 0.0, 0.13}
\usepackage[colorlinks=true,allcolors=burgundy]{hyperref}%
\usepackage[labelfont=bf,justification=centering]{caption}
\usepackage{subcaption}
\usepackage{floatrow} 
\usepackage{tabularx,booktabs,graphicx}
\usepackage{placeins,graphicx,array,makecell,colortbl,multirow,longtable}
\usepackage{float}
\floatstyle{plaintop}
\restylefloat{table}
\usepackage{tikz}
\usetikzlibrary{decorations.pathreplacing}
\usepackage{pgfgantt}
\ganttset{group/.append style={orange},
milestone/.append style={red},
progress label node anchor/.append style={text=red}}

\definecolor{hhsblue}{RGB}{45, 91, 166}

\definecolor{kugreen}{RGB}{50,93,61}
\definecolor{kugreenlys}{RGB}{132,158,139}
\definecolor{kugreenlyslys}{RGB}{173,190,177}
\definecolor{kugreenlyslyslys}{RGB}{214,223,216}

\definecolor{kured}{RGB}{144,26,30}
\definecolor{kuredlys}{RGB}{199,48,40}
\definecolor{kuredlyslys}{RGB}{219,59,10}

\newtheorem{theorem}{Theorem}
\newtheorem{lemma}{Lemma}
\newtheorem{proposition}{Proposition}

\newtheorem{definition}{Definition}
\newtheorem{example}{Example}

\title{\vspace{-95pt} {\fontsize{30}{30}\selectfont{Playing the system: address manipulation and access to schools}}\thanks{\scriptsize{We thank Simon Burgess, Thomas S. Dee,  Ellen Greaves,  Stephanie von Hinke, Ingo Isphording, Edwin Leuven, Emma Tominey, Christine Valente, and Miriam Wüst for helpful comments. We are also grateful for helpful feedback at the  FAIR Virtual Workshop on the Economics of Education in 2021, at  EALE 2021,  at the Workshop on Health, Human Capital and Social Insurance in Bergen in 2022, at IWAEE 2021,  as well as at seminars at the University of Bristol, at the University of Copenhagen, at Lund University, at Aarhus University, at Stockholm University, at UC3M, at University College Dublin, at the Florida Applied Economics Seminar series, and at the University of Reading.   The Norwegian Research Council supported this research under project no. 275906.  The usual disclaimer applies.}}}
\date{Version: \today{} }
\author{Andreas Bjerre-Nielsen\thanks{\scriptsize{University of Copenhagen. Contact: abn@sodas.ku.dk.} }
\and Lykke Sterll Christensen\thanks{\scriptsize{University of Copenhagen.  Contact: lykke.s.christensen@econ.ku.dk.}}
\and Mikkel Høst Gandil\thanks{\scriptsize{University of Oslo, CESifo.  Contact: mikkel.gandil@econ.uio.no.} }
\and  Hans Henrik Sievertsen\thanks{\scriptsize{VIVE, University of Bristol, IZA. Contact: hhs@vive.dk.}}
}
\makeatletter
\AddToHook{env/tabularx/begin}{\let\input\@@input}
\makeatother

\begin{document}
\maketitle
\begin{center}

\begin{minipage}{1\linewidth}
\begin{abstract}
\vspace{-20pt}
\singlespacing

\fontsize{10}{12}\selectfont
Strategic incentives may lead to inefficient and unequal provision of public services.  A prominent example is school admissions. Existing research shows that applicants "play the system" by  submitting school rankings strategically. We investigate whether applicants also play the system by manipulating their eligibility at schools. We analyze this applicant deception in a theoretical model and provide testable predictions for commonly-used admission procedures. We confirm these model predictions empirically by analyzing the implementation of two reforms. First, we find that the introduction of a residence-based school-admission criterion in Denmark caused address changes to increase by more than 100\% before the high-school application deadline. This increase occurred only in areas where the incentive to manipulate is high-powered. Second, to assess whether this behavior reflects actual address changes, we study a second reform that required applicants to provide additional proof of place of residence to approve an address change. The second reform significantly reduced address changes around the school application deadline, suggesting that the observed increase in address changes mainly reflects manipulation. The manipulation is driven by applicants from more affluent households and their behavior affects non-manipulating applicants. Counter-factual simulations show that among students not enrolling in their first listed school, more than 25\% would have been offered a place in the absence of address manipulation and their peer GPA is 0.2SD lower due to the manipulative behavior of other applicants.  Our findings show that popular school choice systems give applicants the incentive to play the system with real implications for non-strategic applicants. \vspace{4pt}\\

\end{abstract}
\end{minipage}
\end{center}
\section{Introduction}
Assigning students to schools is crucial in education policy as it can greatly affect labor market outcomes and the development of social skills and preferences \citep{chetty2011does,sacerdote, rao}. A challenge is to assign students in a way that promotes fairness and efficiency. However, many admission systems allow sophisticated applicants to rank schools strategically, improving their outcomes at the expense of more eligible applicants. In such a case, the rejected applicants are said to have justified envy toward accepted applicants. Modern admission systems aim to eliminate strategic ranking and justified envy, creating a level playing field in the application process \citep[see e.g.,][]{abdulkadirouglu2005new,abdulkadirouglu2005boston,agarwal2018demand,calsamiglia2020structural,haan2023performance}.

Strategic behavior is, however, not limited to just ranking schools. Applicants may try to manipulate ostensibly immutable traits like test scores and place of residence. If an applicant gains access to a school due to manipulation, it may cause the rejection of otherwise more eligible students. In other words, manipulation may cause another form of justified envy, which motivates a broader investigation of strategic behavior around the admission process. 
While anecdotal evidence suggests that manipulation of eligibility is common worldwide \citep[see e.g.,][for the US, England, and India]{nyt,bbc,forbes2018school, guard, toi}, there is little rigorous evidence on the existence of manipulation and the consequences for strategic and non-strategic applicants.

In this paper, we present evidence that strategic applicants manipulate their school-specific eligibility, with direct implications for non-strategic applicants. We develop a theoretical model that incorporates non-preference manipulation into school choice and use it to demonstrate that popular school choice mechanisms incentivize applicants to manipulate their eligibility scores, leading to justified envy. In other words, applicants have an incentive to manipulate, for example, their address of residence, to improve their admission chances, and this behavior has implications for non-manipulative applicants who might lose their school seats due to this behavior. 
We empirically test the model-predicted behavior in the context of Danish high schools exploiting an admission reform that placed greater weight on the address of residence as an admission criterion. This admission criterion is one of the most commonly-used around the world.\footnote{See for example news stories from the US, England, and Germany \citep{forbes2018school,guard,spiegel}. Moreover, \citet{burgessschool} show that 88 percent of schools in England use some form of residence-based admission criterion.} The reform introduced stronger incentives to manipulate reported residence in some geographic regions while leaving incentives unchanged in others.     Using full population administrative records of high school applicants in a difference-in-differences framework, we find that the change in incentives led to a 100 percent increase in address changes around the application deadline. The additional moves are predominantly clustered in the month prior to the application deadline. We observe no clustering prior to the reform, which strongly suggests strategic behavior. We show that the increased clustering is due to moves, which would likely not have occurred in the absence of the incentive to manipulate.  We find no evidence that applicants respond by moving earlier than they otherwise would have. 

We observe a strong social gradient in the response to the admission reform. The likelihood of changing address increases by over 160 percent for applicants from the upper half of the local income distribution. The response is driven by applicants living farther away from high schools and those applying to the most popular ones. Among applicants who do not enroll in their first-choice school, 26 percent experience justified envy due to address manipulation. 

We find great variation in the behavioral response. On average the application incentives increase address changes by 0.6 percentage points (or 0.9 percentage points  for applicants from the upper half of the local income distribution) from a base of 0.6 percent. However, depending on the distance to popular high schools, the increase is up to more than five percentage points and our results suggest that in some high schools, more than one out of ten students in a cohort changed their address in response to the application incentives. 

To understand how this behavior is affected by other policies, we exploit a second policy aimed at mitigating address manipulation. The policy was implemented in the municipality of Copenhagen, which rejected address changes by minors without further proof of residence. The second reform almost eliminated the initial response in the municipality, demonstrating that the response was purely due to manipulation and that a simple, though possibly costly, policy can effectively eliminate strategic behavior.

As applicant strategies affect other applicants, the effect of manipulation on admission to high schools cannot be assessed in standard regression frameworks. We, therefore, reconstruct the assignment mechanism to simulate counterfactual school allocations. We use our estimates of behavioral responses to classify whether moves are manipulative and simulate a policy change that seeks to prevent fraudulent moves. The simulation results show that manipulation harms students from lower-income families, pushing them out of their first priority and into schools where the average peer GPA is 0.2SD lower than in their first priority. Manipulation thus has implications for equity, and policy-makers must weigh the possible welfare loss against the administrative costs of closer scrutiny of address changes.

Finally, we also surveyed applicants to validate our findings based on administrative records. Applicants from highly-incentivized areas are well-informed about the assignment mechanism, and their perception of the costs of manipulation is low. 13 percent of respondents reported considering changing their address to improve their admission chances, and 4 percent reported actually doing so. The survey thus corroborates our finding that manipulation is the driving force behind the change in moving patterns around the reform.

The Danish school choice setting is an excellent setting for studying strategic behavior for at least four  reasons. Firstly, as described above, Danish schools use the address of residence as a criterion. This is one of the most popular criteria to prioritize among students. Secondly, together the two reforms reflect the development seen in many other countries where policymakers first introduced residence-based school assignment mechanisms and then later introduced policies to make it harder to manipulate the address of residence. For example, in England, authorities  started to collaborate across departments and with neighboring authorities to detect false addresses \citep{guard2008}. Thirdly, we believe that the costs of changing addresses in Denmark are neither exceptionally low nor high compared to other countries. While applicants initially could change their address online, they had to change their address in the official centralized records before applying to a school. The address would therefore be the basis for all interactions with the public sector. A fraudulent address may cause changes in taxes and transfers and is punishable by a fine. The address change is therefore more costly than simply listing a fraudulent address on school application forms. On the other hand, if an application for an address change is rejected, no other sanctions are imposed on the applicant.
Finally, the two policy changes combined with the detailed population-wide Danish data allow us to isolate the strategic responses and identify who plays the system and how it affects the other applicants. Given these four reasons, we are confident that our findings have the potential to guide and inform policymakers in various countries and sectors.

This paper contributes to the  literature on manipulability and robustness in matching mechanisms \citep[see e.g.,][]{roth1982economics,abdulkadirouglu2005new,abdulkadirouglu2005boston,calsamiglia2020structural,aygun2021college,grigoryan2022transparency, haan2023performance}. We broaden the strategic game by considering robustness in terms of manipulating eligibility scores. We demonstrate how the truthful revelation of preferences captures only a subset of the potential manipulability in school choice. While an emerging literature strand focuses on behavioral aspects of preference manipulation \citep[see, e.g.,][]{fack2019beyond,rees2023behavioral,chrisander2023students}, the existing evidence on other kinds of manipulation is lacking. \cite{cullen2013jockeying} show that college admissions that favor a share of top-ranked students at all high schools led strategic applicants to enroll in less competitive schools. \cite{aygun2021college} show strategic behavior in the self-reported racial background under affirmative action quotas. \citet{zednik} finds that families in Austria change address more frequently in the relevant months for school admission. We show that the manipulation changes the school assignment with distributive consequences. It is, therefore, essential to consider the appropriate scope for the game when implementing mechanisms in the real world.

Our contribution to the literature on school choice and market design is to develop a general model framework that incorporates manipulation beyond preference misrepresentation. Using this framework, we empirically test how changes in admission criteria affect strategic behavior and how policies can curb such behavior. We find that manipulation of eligibility scores has important externalities and shrinks access to schools for other students, especially those from lower-income families. Evidence from other settings shows that access to selected peers for disadvantaged students is important for educational mobility \citep{cattan2022first}. Our findings demonstrate that the robustness of matching mechanisms depends not only on applicants' strategic incentives in their submitted rank order lists but also on their ability to manipulate eligibility criteria such as the address of residence.

We also contribute to the literature on school choice in relation to residential sorting \citep[see e.g., ][]{nechyba2000mobility,epple2003neighborhood, ferreyra2007estimating,caetano2019neighborhood}. An important distinction in our setting is that we are able to identify short-run responses to incentives in terms of manipulation. As the responses are driven by applicants not actually changing the place where they live, but only their official address, one can argue that this channel may reduce the school-driven residential sorting. It does so by providing access for applicants far from the school, but the mechanism is non-transparent, illegal, and hurts applicants from low-income households.  

This paper is organized as follows. Section \ref{sec:theory} presents the theoretical model. Section \ref{sec:inst} presents the institutional setting. Section \ref{sec:rf} presents the empirical results. Section \ref{sec:conclusion} concludes.

\section{A theoretical model of deception}\label{sec:theory}
We here present our extended school choice framework for understanding deception in two-sided matching. This framework allows us to examine why and how agents engage in deceptive behavior and provides predictions for our empirical analysis. 
We consider a two-stage school choice situation. The second stage corresponds to the standard school choice framework consisting of students who apply to schools through a centralized matching mechanism \citep{abdulkadirouglu2003school}. However, we depart from the standard framework by introducing a first stage where each student decides whether or not to engage in deception to manipulate school priorities. 

Consider an economy that consists of a finite number of students and schools, denoted respectively by the sets, $\mathcal{I}$ and $\mathcal{S}$. Each school $s\in\mathcal{S}$ has a finite capacity for enrollment $C_s$. 
Let a matching $\mu$ be a mapping $\mu: \mathcal{I} \rightarrow \mathcal{S} \cup \emptyset$ where every student is assigned at most one school and each school is assigned no more than their capacity.
To incorporate deception into the model framework, we allow the admission priority that school $s$ gives to student $i$ to be a function of the student's preparatory action,
$a_i\in A_i$, chosen in the first stage. 
We let the priority score be a function of two components: the \textit{manipulable} one $p^{manip}_{i,s}(a_i)$ is a function of the preparatory action, which outputs values in the range $[0,1]$;  the \textit{non-manipulable} one, $p^{exog}_{i,s}\in[0,1]$, which is unaffected by the preparatory action. 
We do not impose structure on the set of actions, but only assume that it has at least two options, $|A_i|\ge2$. This allows the set to encompass a wide range of scenarios, such as the effort required to cheat on an exam or the feasible addresses to which one may relocate.
We define a null action, $a_i^0$, which may be interpreted as, e.g., not engaging in exam cheating or not changing one's address.
We construct the binary measure of student deception as the indicator function $d_i(a_i)$ that equals one if $a_i\neq a_i^0$ and zero otherwise. 
We denote the level of manipulability by the parameter $w\in[0,1]$ which is used to weigh the two school priority components:
\begin{equation}
    p_{i,s}(a_i) = w\cdot p^{manip}_{i,s}(a_i)+(1-w)\cdot p^{exog}_{i,s} \label{eq:priority_manipulate}
\end{equation}



In the final stage of school choice, students report their rank order list over schools, $L_i$. Schools observe the realized priorities conditional on the preparatory actions of students but we assume that the actions are private information to the students. Students are assigned to schools using a matching mechanism. The mechanism specifies a message space in which each student can submit his/her preferences over schools. Students are assigned to schools based on a matching procedure, which uses students' rank order lists as well as schools' capacities and priorities (higher priorities first) as input. We consider Student-Proposing Deferred Acceptance (DA) or Immediate Acceptance (IA) as mechanisms, see \cite{abdulkadirouglu2003school} for details.\footnote{Existing literature sometimes refers to Immediate Acceptance as the Boston Mechanism.}

Student $i$ receives utility $v_{i,s}\in \mathbb{R}$ when matched to school $s$. We assume that unassigned students receive a utility of zero and that each student's ordinal preferences are strict over schools $\mathcal{S}$. When an agent $i$ deceives then she incurs a strictly positive cost $\gamma$. The cost parameter, which we assume is the same for all agents, may capture inconvenience or loss from penalties if caught in the deception act. We assume that students only deceive if they strictly benefit in terms of utility.\footnote{In technical terms, this means that when a student is indifferent between admission at one school $s$ after deception vs. entry at another $s'$ without deception ($v_{i,s}-\gamma=v_{i,s'}$) then the agent always prefers admission at $s'$ without deception.}

\paragraph{Strategic concepts and properties}
To depict how this model works, we now provide an example of how deception can arise and what the consequences are. 

\begin{example}\label{example:strategic}
Suppose there are two students and a single school with one seat. Both students get a utility of 1 from being admitted to the school, which exceeds the utility of being unmatched and receiving a utility of zero. The seat is allocated using DA (or IA, which is identical in this setting). Student 1 has a lower non-manipulable priority score than student 2. If deception is not feasible (i.e. $w=0$) then student 2 gains admission and the total welfare is $1$.

Assume now instead that deception is possible and exclusively determines priority ($w=1$). Suppose also that if student 1 uses deception it has a higher priority than student 2, irrespective of student 2's preparatory action. Finally, assume that the cost of deception ($\gamma$) is lower than the utility of admission ($=1$).  Naturally, student 1 deceives to gain admission and thereby increases its utility, and thus student 2 is rejected. The total welfare for the students is $1-\gamma$. \end{example}

The above example illustrates how students may benefit from deception in the admissions process, potentially harming other students and their well-being.

We proceed to extend strategic measures from the literature and analyze their properties, see Appendix~\ref{app:proofs_strategic} for proofs.  A mechanism is \textbf{strategy-proof} if for every student and for all of her preparatory actions, and for all strategies of other students, it is optimal for her to report her ordinal preferences truthfully. A mechanism is \textbf{deception-proof} if for every student and for all strategies of other students, it is optimal to not deceive. 

\begin{proposition}\label{proposition:manipulability}
DA is strategy-proof but IA is not, however, neither is deception-proof.
\end{proposition}

The proposition above demonstrates how the truthful revelation of preferences captures only a subset of the potential manipulability. That is, although DA is strategy-proof in terms of preferences, it is not deception-proof as students can use deception.

A mechanism eliminates \textbf{justified envy} if, for every preference and every action $a_i$ with its realized priority profile, i.e. $p_{i,s}(a_i)$, there exists no pair of a student and a school such that the student prefers the school over her assignment and the school either has a vacant seat or the student's priority is lower than that from another student matched to $s$.
We amend the standard concept of justified envy by fixing priority structures so they are not affected by student deception. We say that a mechanism eliminates \textbf{invariant justified envy} when we instead compare students' priority under the assumption that $w=0$. In other words, this requires that no students prefer a school where their non-manipulable priority component exceeds that of students admitted there.

\begin{proposition}\label{proposition:envy_free}
DA eliminates justified envy but not invariant justified envy.
\end{proposition}

The above result demonstrates that the elimination of justified envy to the non-manipulable priority score is not possible. Intuitively, this impossibility follows from the fact that invariant justified envy can be violated whenever there is an incentive to deceive, which arises naturally whenever the costs of deception are low and the manipulability of priority is high.

\paragraph{Optimal deception}
We proceed with an analysis of optimal deception behavior of an agent. To define the optimal deception, we introduce matching cutoffs as they allow us to keep other parts of students' behavior fixed. 
For a given matching $\mu$, we define \textbf{cutoffs} as the minimal priority scores $P_\mu\in \mathbb{R}^{S}$ required for admission at each school. When the number of students admitted at a school is lower than its number of seats, then any student can gain admission there, which we denote by $P_s=-\infty$. This implies that there is a single cutoff for each school, which determines the entry requirements for every student. Note that, unlike in a standard matching framework, here, cutoffs are also determined by deception. 

Optimal deception is the level of deception that together with a rank order list maximizes a student's utility when the school cutoffs and level of manipulability are considered exogenous (from the student's perspective); 
a formal definition is found in Appendix~\ref{app:supply_demand} along with proof of the following statement. 

\begin{proposition}
\label{claim:monotone_deception}
Let cutoffs be fixed; the optimal deception is i) non-decreasing in the level of manipulability ($w$) and ii) non-increasing in the cost of deception ($\gamma$).
\end{proposition}

The above result shows that optimal deception is weakly monotone in the level of manipulability, which provides a foundation for our subsequent empirical analysis. 
We analyze the full equilibrium behavior in large matching markets in Appendix~\ref{app:large_economy}, where we also construct the aggregate deception as a function of the level of manipulability captured by Equation \eqref{eq:aggregate_deception_manipulability}. We provide conditions for the effect of manipulability to be positive, which is captured by the sign of Equation \eqref{eq:aggregate_deception_manipulability}. Higher levels of manipulability have a direct positive effect. However, there is also a general equilibrium effect that operates through changes to other schools’ cutoffs which is ambiguous. 

\paragraph{Testable predictions}
We conclude our theory section by using our model framework to provide testable predictions. We operationalize our empirical framework as follows. Schools have addresses, which we interpret as spatial coordinates $a_s\in\mathbb{R}^2$. Each student $i$ chooses an address among its choice set, which are also interpreted as coordinates  $A_i\subseteq \mathbb{R}^2$. 
We let the manipulable part of the priority score equal either the negative Euclidian distance between students and schools, i.e., $p^{manip}_{i,s}(a_i) =-\lVert a_i-a_s\rVert_2$ or negative transportation time between their addresses, such that lower distance/transportation time gives higher priority score.

In this spatial context, deception consists in moving away, i.e., getting another address. As a direct corollary of Propositions~\ref{claim:monotone_deception} and \ref{claim:marginal_deception_agg}, we predict that a higher level of manipulability ($w$) leads to more students relocating and that a higher cost of manipulation ($\gamma$) leads to fewer students relocating. In what follows, we begin by analyzing the first prediction empirically by focusing on the introduction of a new admissions system in Denmark. The new system was based on the IA matching mechanism with distance-based priorities and thus we interpret the change in admission as an increase in $w$, see details in Section~\ref{sec:inst}. To test our prediction, we analyze whether or not students' moved away before applying to high school and we outline a quasi-experimental approach to causally measure this effect. Subsequently, we investigate a reform specific to the municipality of Copenhagen that increased the documentation required when children change their addresses. This reform can be interpreted both as a reduction of the level of manipulability ($w$) or an increase in the manipulation costs ($\gamma$). In either case, we expect the reform to lower the number of students relocating.

\section{Institutional background}
\label{sec:inst} 
To causally determine whether deception behavior increases as the weight on the manipulable component of the priority score increases, we use policy reforms in Denmark and individual-level data from administrative records. In this section, we outline the institutional context. We first describe the general education system, including the high school admissions process. Then, we discuss two policy changes that affected the weight on the manipulable component of the priority score.

\subsection{The Danish school system}
\subsubsection{Educational paths}
In Denmark,  compulsory schooling starts in August of the calendar year a child turns six years old and ends after ten years. After completing ten years of schooling, about 55\% of a cohort continues to high school (either in the general academic or technical track), 25\% pursue vocational education and training, and the remaining 20\% either join the workforce or explore other educational options. This study focuses on the most popular high school programs: general academic high schools (DK:STX), which offer a three-year program that provides direct access to university education and other post-secondary programs. The general academic track has multiple sub-tracks with specialized focuses that vary across high schools. Most academic high schools also offer a two-year program (DK:HF) for older students.

During the period of our study, high schools are self-governed institutions. They receive central funding through a yearly grant and a per-pupil amount and are free to allocate these funds as they see fit.  High schools are free to prioritize the funds given. In the period, high schools were more or less able to set capacity themselves and anecdotal evidence suggests heavy competition between high schools in attracting students.

\subsubsection{The high school application process}
Potential students submit their preferences over a maximum of five high schools by March 1 for enrollment in August of the same year. Applicants are matched to schools using Immediate Acceptance. The use of the Immediate Acceptance mechanism implies that all students are tentatively assigned to their first-priority high school in the first round. In the case, where the number of applicants exceeds the capacity of a high school, the students are ranked by an eligibility score which we describe in detail below.  The high school application process is governed by an online platform called “optagelse.dk”. On this platform, background information on the applicants is retrieved from the centralized population registry (CPR) and unchangeable by the applicants. The background information includes information on the applicants’ home addresses and their parents.

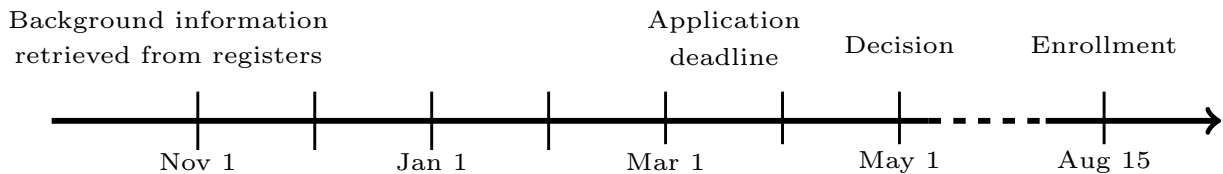
\begin{figure}[t!]
   \resizebox{\linewidth}{!}{
  \begin{tikzpicture}[node distance=1cm]
   \draw  [line width=.25mm] (0.25,1.25) -- (0.25,1.75);
   \draw  [line width=.25mm] (1.25,1.25) -- (1.25,1.75);
   \draw  [line width=.25mm] (2.25,1.25) -- (2.25,1.75);
   \draw  [line width=.25mm] (3.25,1.25) -- (3.25,1.75);
   \draw  [line width=.25mm] (4.25,1.3) -- (4.25,1.75);
   \draw  [line width=.25mm] (5.25,1.3) -- (5.25,1.75);
   \draw  [line width=.25mm] (6.25,1.3) -- (6.25,1.75);
   \draw  [line width=.25mm] (8,1.3) -- (8,1.75);
   \draw  [line width=.5mm ] (-1,1.5) -- (6.5,1.5);
   \draw  [line width=.5mm,dashed] (6.50,1.50) -- (7.5,1.50);
   \draw  [line width=.5mm,->] (7.5,1.50) -- (9,1.5);
   \node at (0.25,1.15) {\tiny Nov 1};
   \node at (2.25,1.15) {\tiny Jan 1};
   \node at (4.25,1.15) {\tiny Mar 1};
   \node at (6.25,1.15) {\tiny May 1};
   
   \node at (8,1.15) {\tiny Aug 15};
   
    \node at (0.00,2.35) {\tiny Background information};
    \node at (0.00,2.05) {\tiny retrieved from registers};
    \node at (4.75,2.35) {\tiny Application};
    \node at (4.75,2.05) {\tiny deadline};
    \node at (6.25,2.15){\tiny Decision};
    \node at (8,2.15){\tiny Enrollment};

   \end{tikzpicture}}
   \caption{Timeline of Application Process}
   \label{fig_timeline}
 \end{figure}

Figure \ref{fig_timeline} outlines the application process. The process starts in October the year before enrollment. At this point in time and until the application deadline, the current schools of all applicants must retrieve the applicants' background information from the population register such that the information used in the application formula is up to date. When the application window opens in January, applicants can log into the system and check the correctness of their background information. In case of incorrect information, applicants must contact their current school and get them to update the data.

When applicants have registered their ranked ordered preferences over high schools and submitted their application, a parent or guardian must sign the application electronically. The application window closes on March 1 and the allocation committee takes care of the formal allocation of applicants. 
In May or June, most applicants receive information on which school, they have been allocated to before actual enrollment in August.

\subsection{Two policy changes and model predictions}
We will now describe two policy changes that impacted the high school admissions process. The first, implemented in 2012, directly altered the construction of the high school priority score. The second, implemented in one municipality in 2018, did not directly affect the admissions process but made the manipulable component less manipulable and increased the cost of deception.
In the subsequent empirical analysis, we will mainly focus on the first policy change as it is almost identical to a change in the theoretical parameter that governs the level of manipulability, $w$, which we presented in section \ref{sec:theory}. The second policy change only affected one municipality and was made in response to the first one and is, therefore, less suitable for empirical identification as it only indirectly affected the high school admission process. However, the second reform allows us to validate the conclusions drawn from the evaluation of the first reform. 

\subsubsection{Reform I: The 2012 high school priority reform}

If the number of applicants with first preference to a school exceeds the school's capacity, the allocation of the students is decided by the regional assignment committee. Up until 2011, the assignment committee had vague instructions on how to prioritize between students, but they were asked to \emph{also} take the distance between the school and the individual's residence into  \emph{consideration} (see Appendix Figure \ref{fig:pre2012law}). Following increasing student numbers and public requests by high school principals that the vague rules led to unfair battles between schools, the rules were changed in 2012. Most importantly, as of 2012, the high school assignment committees had to assign students based on travel time with public transport or walking time (see Appendix Figure \ref{fig:post2012law}). In 2017, the assignment policy was changed from travel time to travel distance to ensure that applicants living closest to a given high school had the highest priority \citep{berlingsk2016nyregel}. The measures of travel time or distance are constructed by accessing the information on the individual students' addresses from a central registry\footnote{Before 2016, the assignment committee was instructed to use the publicly accessible homepage "www.rejseplanen.dk" to calculate travel time. After 2016, the travel distances were calculated using google maps.}. In practice, the process of changing an address in Denmark can be done instantaneously online using individual secure login systems. In particular cases, the municipal administration can halt the registration to review it. Changes to these review processes were the scope of the second policy change, see below.

The vague descriptions for giving students priorities before 2012, meant that applicants were unable to predict what their priority would be and therefore also made it very difficult for them to manipulate their scores. 
Since 2012, geographic distances were the only determinant of students' priority scores. Given how straightforward it is to change one's home address in Denmark, this implies that the level of manipulability approaches its highest level ($w\rightarrow 1$) due to this reform. Consequently, students who can re-register as living at other locations, e.g. in the home of family or friends of their parents, or by renting a place, can improve admission chances at specific schools if at least one of these locations is nearer to a school they like than their current home. Drawing on Propositions~\ref{claim:monotone_deception} and \ref{claim:marginal_deception_agg}, our first hypothesis is thus that the first reform increases deception.

\subsubsection{Reform II: The 2018 address manipulation reform}
In principle, any municipality must approve formal address changes made by minors when they move without a parent and do not move to a place where a parent already lives. However, during the period considered, municipalities varied in the amount of documentation required to approve the address change.

Since 2018, the municipality of Copenhagen has increased its efforts to combat address manipulation. This includes sending a letter to the minor’s parents or legal guardians when an address change is requested. The letter requires documentation that convincingly establishes that the move is genuine.\footnote{This could include copies of the lease, proof of moving expenses, commuting costs between the new address and elementary school, expenses for refurbishing a new room, etc.}
While the case is pending, the address change is put on hold and not officially registered. This means that if the purpose of the address change was to increase the chances of being admitted to a particular high school, the effort may be fruitless if the case is not settled before March 1st. If the parents or legal guardians fail to convincingly establish that the move is genuine, the municipality rejects the address change. The decision is communicated by letter with a review of the process and an explanation of the decision. Neither the parents, legal guardians nor the minor will be fined or otherwise penalized and they could try to register a move again.

In terms of our model framework, the second policy reform implied that students engaged in deception involving addresses in Copenhagen Municipality now face a higher deception cost (increased $\gamma$) and face a risk that their action is reversed to the null-action (corresponding to a lower weight on the manipulable part ($w$)). Drawing on Propositions~\ref{claim:monotone_deception} and \ref{claim:marginal_deception_agg}, our second hypothesis is thus that the second reform decreases deception.

\section{Empirical evidence}
 \label{sec:rf}
In this section, we assess the reforms' impact and test our model's predictions. After detailing our data and descriptive statistics, we describe our approach to identify manipulation effects. We then present our estimates and evaluate the distributive consequences of manipulation. Finally, we present results from a survey to applicants that support and detail our findings.

 \subsection{Data}
\subsubsection{Data sources and sample selection}
Our sample consists of the universe of students in the last year of compulsory schooling (grade 9). We merge this sample with a number of background registries to obtain information about students' date of birth, gender, daily level address changes, high school applications
We also add information on the students' parental income (from tax records) and educational attainment (from education registries). From education registries, we also get information on the individuals' middle school GPA (defined as attainment in the last year of compulsory schooling, before entering high school).

For the main analysis, we use the 535,114 students finishing compulsory schooling in the years 2009 to 2017, which covers all available data up until the second reform. We further restrict the sample to the 147,903 students who applied to a high school during the last year of compulsory schooling, which has no impact on results.\footnote{We show that our conclusions are not affected by considering the full cohort who finish compulsory schooling}. %
In the analysis of the effects of the second reform, we consider data up until 2020, which is the last year where we observe all variables across the registries. 

Furthermore, we have collected survey data from applicants to high schools in the years 2020 and 2021. In these surveys, we  asked students about their true ranked order list of high schools, if they were guaranteed access, and whether they have considered changing their address to gain access to their most preferred high school. The response rate was about 20 percent in 2020 and slightly lower in 2021, where it was 15 percent We return to the survey results at the end of the empirical setting to validate results and shed light on the considerations behind the behavior we identify.

\subsubsection{Variable definitions}
Our main outcome of interest is address changes linked to the high school application process. This binary variable is set to 1 if the student relocated between the start of the final pre-high school year (October 1) and the high school application deadline (March 1), and zero if no change occurred.    We explore within-year variation graphically and condense this data into two key variables for regression analyses: the original address, and an address change indicator, which is 1 if a change occurred during the specified period. As few students change their address several times, we consider only the first address change.

Using data from income registries, we create a measure of average parental disposable income, where the average is computed across the observed parents in the data. This income measure is measured after taxes and transfers and converted to the 2015 price level using the consumer price index. Using data from the education registry we create a measure of the average years of schooling of the observed parents. We also create a measure of  middle school GPA as the grade point average based on both exam results and teacher evaluations in the last year of compulsory schooling. We standardize this measure to have a mean of zero and a standard deviation of one within each cohort. 

We define over-subscribed high schools based on how often a school is listed first in applications, compared to how many students eventually get enrolled. A high school that is listed as number one more often than the number of students they enroll is defined as over-subscribed. 

\subsubsection{Descriptive patterns}

Table \ref{tab:sumstat} provides summary statistics for key variables in the full sample of students who completed compulsory schooling from 2009 to 2017, as well as for the subsample of students who applied for high school enrollment. The table shows that girls are slightly more likely to apply for high school and that high school applicants are, on average, children of parents who have completed more years of schooling and have higher incomes than the general population.

\begin{table}[h!]
    \caption{Summary statistics}
    \label{tab:sumstat}
    \begin{tabularx}{1\linewidth}{cXcc|ccccc}
        \toprule
    &&\multicolumn{2}{c}{--- All --- }&\multicolumn{3}{c}{Means for HS applicants }\\
    &&\multirow{2}{*}{Mean}&\multirow{2}{*}{SD}&\multirow{2}{*}{All}&\multicolumn{2}{c}{Over-subscribed }\\
    &&&&&No&Yes\\
            \midrule
        \multicolumn{6}{l}{A. Individual characteristics }\\
        &Female&  0.49&  0.50&  0.60&  0.61&  0.60\\
&Age& 15.83&  0.45& 15.78& 15.79& 15.77\\
&Parental schooling (years)& 13.90&  2.69& 14.65& 14.38& 14.93\\
&Parental income (1000 USD)& 41.02& 54.17& 47.34& 44.63& 50.23\\
\\
        \multicolumn{6}{l}{B. High school application behavior }\\
       &Applied to high school&  0.28&  0.45&  1.00&  1.00&  1.00\\
&Applied to oversubscribed high school&  0.49&  0.50&  0.49&  0.00&  1.00\\
&Enrolled in high school&  0.25&  0.43&  0.88&  0.90&  0.87\\
&Enrolled in oversubscribed high school&  0.43&  0.49&  0.44&  0.00&  0.90\\
&Enrolled in 1st priority high school&  0.83&  0.23&  0.83&  0.89&  0.78\\
\midrule
&Observations& 534,114& & 147,903&  74,737&  72,273\\
\\[-12pt]
        
        \bottomrule
        \end{tabularx}
        \begin{minipage}{1\linewidth}
                        \footnotesize
        Notes: Parental schooling is the average among observed biological parents.  Parental income is the average income in 1000 EUR (2015 level), after taxes and transfers averaged across the observed biological parents. Both parental schooling and income are measured in the year of high school entry. A high school is defined as over-subscribed if more students listed the high school first in their application than the total number of students enrolled.
        \end{minipage}
\end{table}

For the main analysis, we consider the 28 percent who apply to high school during their last year of compulsory schooling. Looking at their application behavior, Table \ref{tab:sumstat} shows that 49 percent of applicants apply to over-subscribed high schools, 88 percent of the students who apply also enroll, and 83 percent of the students enroll in the high school they listed first in their submitted rank order list. 

The last two columns of Table \ref{tab:sumstat} show that the socioeconomic gradient in applying to any high school also is present in terms of applying for an over-subscribed high school. Applicants to over-subscribed schools have parents who have completed more years of schooling and who higher income compared to those who apply to a school that is not over-subscribed.

\begin{figure}[h!]
\begin{center}
    \includegraphics[trim=2cm 1cm 2cm 1.cm, clip, width=\linewidth]{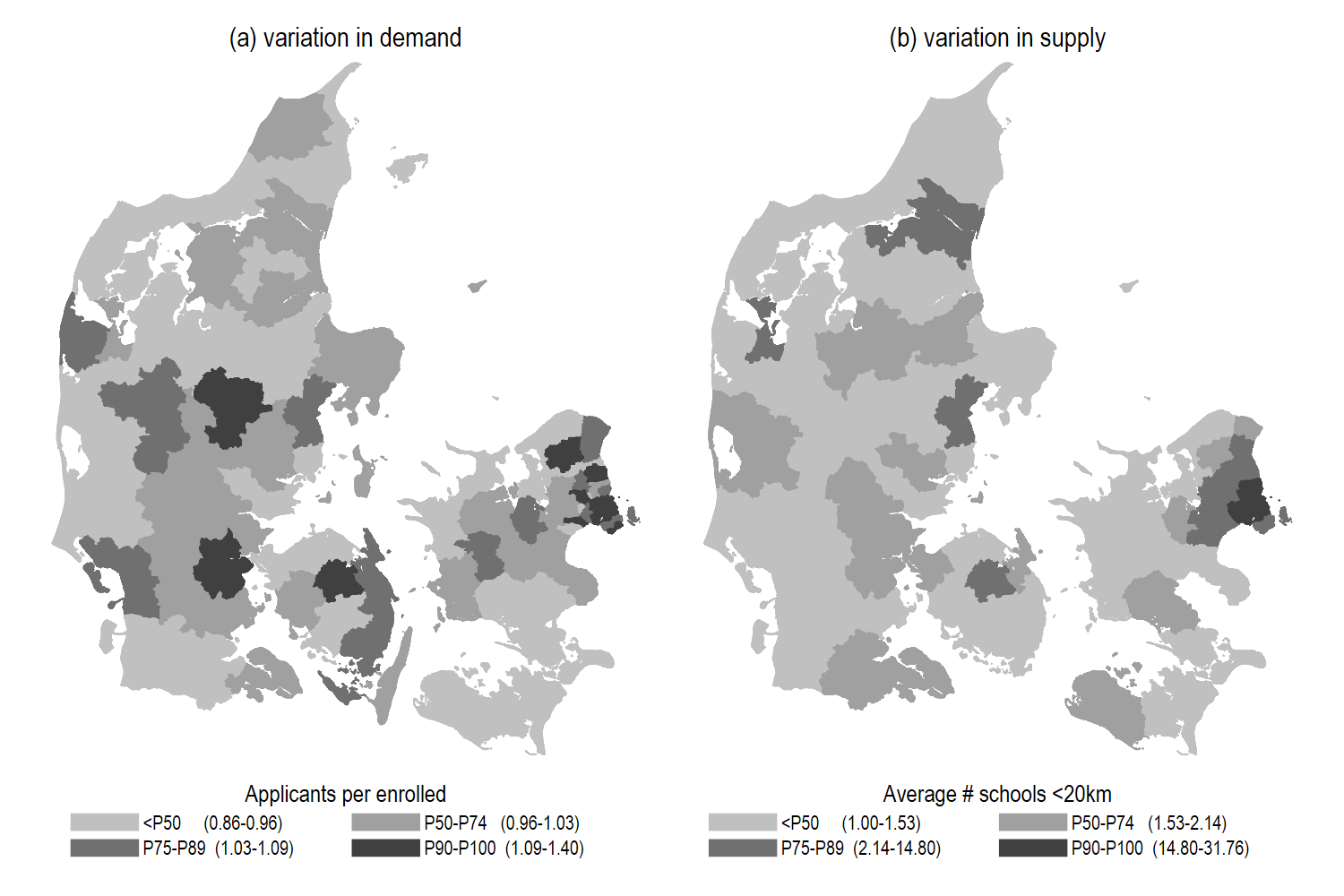}
\caption{Geographic variation in demand and supply of high schools\vspace{8pt} \\ 
\begin{minipage}{1\linewidth}
\footnotesize
Notes: The  demand is measured as the number of students listing a particular school first relative to the number of students this school enrolls. The map then shows the average of this ratio across all students in the municipality. The supply is measured in terms of the number of high schools within 20km distance from the applicants residence at the beginning of the last year of compulsory schooling. The distance is the euclidean distance from the center of the a 1000m$\times$1000m square around the individual's address. Both demand and supply is based on averages for the period 2009 to 2011.
\end{minipage}}
\label{fig:geographvara}
\end{center}

\end{figure}

Figure \ref{fig:geographvara} documents the variation in demand and supply of high schools across Danish municipalities. Figure \ref{fig:geographvara} (a) shows that demand varies considerably across municipalities. Using only data before 2012 we construct the ratio of applicants to enrolled for the high schools to which students apply. In the municipalities in the top 10 percent of the distribution of ratios, students on average applied to a high school that received between 1.09 and 1.40 applicants per enrolled student.  

Figure \ref{fig:geographvara} (b) shows the geographic variation in the supply of high schools in terms of the number of high schools within 20km distance from the applicant's residence at the beginning of the last year of compulsory schooling. Similar to the demand measure, we observe a pattern where supply is largest around the larger cities. For example in the top 10 percent of the municipalities in terms of supply of high schools, the applicants on average have access to between 14.80 and 31.76 high schools within 20km. In contrast, in the municipalities below the median, students have access to between 1 and 1.53 high schools. 

\begin{figure}[h!]
\includegraphics[width=1\linewidth]{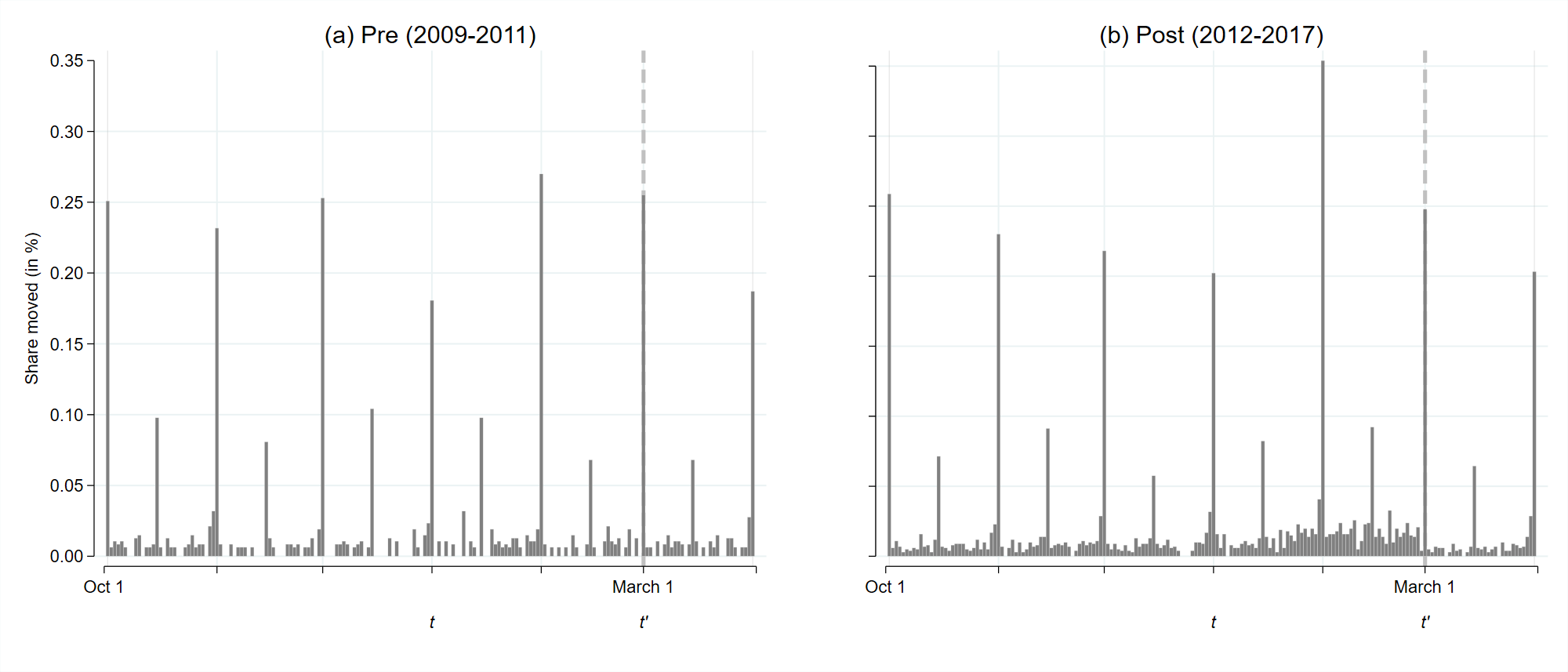}
\caption{Daily address changes  by applicants in the five months up to high-school application deadline on March 1 \vspace{8pt} \\ 
\begin{minipage}{1\linewidth}
\footnotesize
Notes: The figure shows the daily rate of address changes for all students in the last year of compulsory schooling who applied to a high school. The share is calculated relative to the population  on October 1. 
\end{minipage}}
\label{fig:pdfs}
\end{figure}

Before we turn to the empirical analysis it is useful to consider the raw patterns in address changes. In Figure \ref{fig:pdfs} we show the daily address changes in the period from October 1 to April 1 of the last school year before high school enrollment among those who apply for high school. The figures show that both before and after the 2012 reform, there are spikes in address changes on the first of the month and somewhat lower in the middle of the month. We also observe that address changes are relatively rare and peak at about 0.25 percent on a given day. The chart also shows the first evidence of a behavioral change after the 2012 reform where the weight on manipulable component of the priority score, $w$, increased considerably, as we see substantial mass prior to the application deadline on March 1 in the post-reform period shown on the right, but not in the pre 2012  period shown on the left.

\FloatBarrier

In the next subsection we will combine the geographic variation in demand shown in Figure \ref{fig:geographvara} with the change in incentives before and after 2012 to identify the causal effect of application incentives on address changes.

\subsection{Empirical strategy}
\label{subsec:strategy}
The main goal of the empirical analysis is to assess the predictions from section \ref{sec:theory} that an increase in the weight on the manipulable component of the priority score, $w$, will lead to increased manipulation. While the patterns shown in the charts in Figure \ref{fig:pdfs} suggest a behavioral response to the 2012 where $w$ increased, further analyses are required to quantify the response and rule out alternative explanations.

To quantify the increase in manipulation we need to identify the address changes that only happened because of the increase in $w$. We do so in a standard difference-in-differences design where we exploit two sources of variation.

We define students into two groups based on their geographic location. A treated group, $T$, that is affected by the change in $w$ and has the incentive to manipulate the admission score, and a control group, $C$, who either doesn't experience a change in $w$ or who has no incentive to manipulate the priority score. We define the treated group as those living in municipalities where demand for high schools is in the top quartile. Those living in the remaining municipalities constitute the control group.

\begin{figure}[t]
\includegraphics[width=1\linewidth]{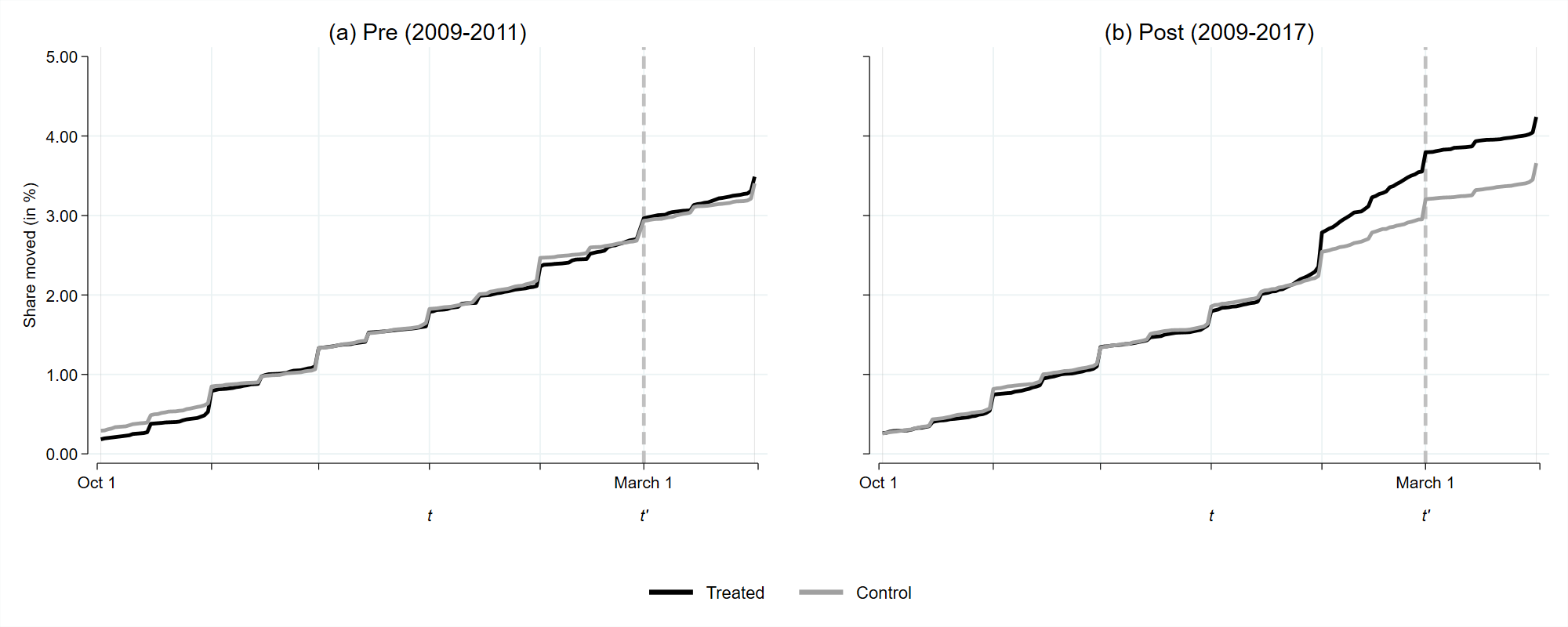}
\caption{Cumulative address changes by applicants in the five months up to high-school application deadline on March 1 \vspace{8pt} \\ 
\begin{minipage}{1\linewidth}
\footnotesize
Notes: This figure plots the cumulative address changes in the control and treatment groups before and after the reform.  We define the treated group as those living in municipalities where demand for high schools is in the top quartile (see Figure \ref{fig:geographvara} (a)). The accumulated share of moves in the treatment group, i.e. applicants living in municipalities with general over-demand of high schools, is shown in black. Correspondingly, the control group, i.e. the remaining municipalities is shown in gray. 
The application deadline is the vertical dashed line. 
\end{minipage}}
\label{fig:addressmanipulation}
\end{figure}

At a given time for each group, we compute the total share of an applicant cohort that moved since October 1st. We focus on the total share at the high school application deadline. Figure \ref{fig:addressmanipulation} shows the two groups before and after the 2012 reform where we pool cohorts prior and post-reform.

Figure \ref{fig:addressmanipulation}(a) shows the share of applicants moving in cohorts who apply prior to the 2012 reform. Here we would expect no difference between the two lines because even though the groups differ in the incentive to manipulate the score, none of the groups have the ability to manipulate the score through changing address. In Figure  \ref{fig:addressmanipulation}(b) both groups have the opportunity to change the score by changing address, but the groups differ considerably in the share that moved. The response is a 0.60 percentage points after the reform, compared to 0.02 percentage points before the reform.  Theoretically, we could consider all address changes since birth, but as the figures show, the behavioral response only starts about one month prior to the deadline, and considering all address changes in the last school year is, therefore, more than sufficient.

 Figures \ref{fig:addressmanipulation}(a) and \ref{fig:addressmanipulation}(b)  illustrate the main idea behind our identifying approach to rule out of alternative explanations. We consider the response in the post reform period, Figure \ref{fig:addressmanipulation}(b), and subtract the same difference from the pre reform period, Figure  Figure \ref{fig:addressmanipulation}(b) in a difference-in-differences approach.

Concretely, we exploit the geographic and time variation in a difference-in-differences approach by estimating the following specification using ordinary least squares:
  \begin{align}
        move_{i}=\alpha+\beta treated_{i}\times post_{i}+\bm{\gamma}'\bm{Y}_{i}+\bm{\theta}'\bm{M}_{i}+\bm{\pi}'\bm{X}_{i}+e_{i}
        \label{eq:1}
    \end{align}

where $move_{i}$ is equal to 1 if the individual changed address between October 1 and March 1 in the last school year before high school. The variable $post_i$ is 1 for students applying to high school after 2011 and 0 otherwise. The variable $treated_i$ is 1 for individuals who lived in a treated municipality at the beginning of their last school year before high school and 0 otherwise. The vectors $\bm{Y}_i$ and $\bm{M}_i$ include indicator variables for respectively the years 2011 to 2017 and for the 98 municipalities, capturing the year fixed effects and the fixed effects for the initial residence.  Finally, the vector $\bm{X}_i$ includes controls for gender, 9th-grade GPA, parental income, and parental education. To allow for arbitrary correlation within the original municipality of residence, we cluster the standard error by the municipality of residence at the beginning of the last year of compulsory schooling.

Our identification strategy exploits that the opportunity to manipulate the admission priority increased for both groups in 2012, but only the treated group had the incentive to do so. If other incentives changed at the same time in the treated regions and not in the control regions, the coefficient $\beta_3$ will also capture the response to these. To assess this assumption, we studied law changes over the same period and also consulted civil servants at ministries and municipalities. To the best of our knowledge, there were no other simultaneous policy changes. However, we also conduct several empirical tests to validate this assumption.

Even in the absence of simultaneous policy changes, diverging time trends in address changes in treated and control groups would also violate our identification, because $\beta_3$ would then also capture the time effect if this trend is not modeled appropriately. We assess this assumption by comparing annual patterns in address changes before and after treatment.

\subsection{Results}\label{subsec:res}
\subsubsection{Evidence of address manipulation}
Table \ref{tab:main_results} shows the regression results from estimating equation \ref{eq:1}. In line with the graphical evidence, we observe a difference-in-differences estimate of 0.006 (i.e. 0.6 percentage points), as shown in column (1) of Table \ref{tab:main_results}.  Compared to the baseline of 0.031, this is an increase of 19 percent. The effect is precisely estimated and significantly different from zero. Columns (2) and (3) of Table \ref{eq:1} show that the response conceals clear heterogeneity. We observe no response by children of parents with a disposable income below the municipality median, but a strong response of 1.1 percentage points, corresponding to an increase of 42 percent relative to the baseline mean, for  children from families with a disposable income above the median. Columns (4) to (7) show that within income groups there is little evidence of heterogeneous responses by middle school GPA.

\begin{table}[h!]
    \caption{Regression results: The effect of high-school application incentives on moving behavior}
    \label{tab:main_results}
    \begin{tabularx}{.95\linewidth}{cXc cc cc ccc}
    \toprule
   &&\multirow{4}{*}{All}&&&\multicolumn{4}{c}{Parental income}\\
    &&&&&\multicolumn{2}{c}{$\leq$p50}&\multicolumn{2}{c}{$>$p50}\\
    &&&\multicolumn{2}{c}{Parental income}&\multicolumn{4}{c}{Middle school GPA}\\
    &&&$\leq$p50&$>$p50&$\leq$p50&$>$p50&$\leq$p50&$>$p50\\
    &&(1)&(2)&(3)&(4)&(5)&(6)&(7)\\
    \midrule
    \multicolumn{7}{l}{\emph{A. All address changes}}\\
   &Post $\times$ treated&       0.005&       0.000&       0.011&      -0.001&       0.002&       0.012&       0.009\\
&                    &     (0.003)&     (0.003)&     (0.005)&     (0.004)&     (0.004)&     (0.005)&     (0.006)\\
&MDV                 &       0.031&       0.037&       0.026&       0.040&       0.032&       0.028&       0.024\\
\\
    \multicolumn{7}{l}{\emph{B. Individual address changes}}\\
    &Post $\times$ treated&       0.006&       0.003&       0.009&       0.003&       0.003&       0.008&       0.010\\
&                    &     (0.002)&     (0.001)&     (0.003)&     (0.001)&     (0.003)&     (0.003)&     (0.004)\\
&MDV                 &       0.006&       0.006&       0.006&       0.006&       0.007&       0.006&       0.007\\
\\
    \multicolumn{7}{l}{\emph{C. Household address changes only}}\\
  &Post $\times$ treated&      -0.000&      -0.003&       0.002&      -0.004&      -0.002&       0.004&       0.000\\
&                    &     (0.002)&     (0.003)&     (0.002)&     (0.004)&     (0.004)&     (0.004)&     (0.003)\\
&MDV                 &       0.025&       0.031&       0.019&       0.034&       0.025&       0.022&       0.018\\
\midrule
\multicolumn{2}{l}{Observations}        &     147,903&      74,155&      73,674&      44,757&      28,751&      29,080&      44,012\\
    \bottomrule
    \end{tabularx}\\
      \begin{minipage}{0.95\linewidth}
        \footnotesize Notes: The table shows the coefficients from estimating equation \eqref{eq:1} using ordinary least squares. The dependent variable takes the value of 1 if the individual changed the address before March 1, and 0 otherwise.  The variable treated is 1 for individuals living in municipalities where demand for high schools is in the top quartile (see Figure \ref{fig:geographvara} (a)) and zero otherwise. The variable post is one in all years after 2011. All regressions include the full set of year and municipality fixed effects, as well as the  full set of controls. The full set of controls includes a variable for the average parental years of schooling, a variable for the average parental disposable income, a variable for the 9th-grade GPA, and an indicator for the biological gender being female. We also include indicator variables for missing parental education, missing parental income, and missing 9th-grade GPA. Panel A. shows results where the dependent variable is 1 if the individual student changed address in the period between October 1 and the last day of February in the school year before high school enrollment, and 0 otherwise. In Panel B only address changes where no other member of the initial household changes address are considered. In Panel C only address changes where at least one other member of the initial household changed address are considered. Columns (2) to (7) show sub-sample regressions based on parental disposable income and middle school GPA, for both measures the median is calculated by cohort and municipality.         MDV is the mean of the dependent variable for untreated individuals.         Standard errors clustered at the municipality of residence at the beginning of the last year before high school enrollment in parenthesis. 
    \end{minipage}
\end{table}

In panel B of Table \ref{tab:main_results} we restrict the address changes to individual changes where no other member of the original household changed address. The response is almost identical to the overall responses reported in panel A, but the baseline is lower at 0.006, which implies a relative increase of 100 percent. When considering only the individual address changes we also find a small and significant response for the low-income group at 0.003 percentage points, corresponding to an increase of 50 percent compared to the baseline mean.  However, for children of high-income families, the response is a 167 percent increase in the likelihood of changing address. Given that the coefficients in panel A and panel B are similar, it is unsurprising that the missing piece, address changes where at least one other member of the household also changed address, shows no response, as shown in panel C of Table \ref{tab:main_results}.\footnote{In Appendix Figure \ref{fig:ab} we show the cumulative address changes by March 1 for respectively the individual and household moves. In line with the regression results, we observe a strong response by individual moves after the reform and no response by household moves. Moreover, the charts show that both individual and household addresses changes in the treatment group follow the pattern of the control group closely in the pre-reform period.}

\subsubsection{Identifying assumptions and robustness}
Before we turn to the mechanisms of how high school applicants manipulate their addresses, we first assess the identifying assumptions. We first assess the assumption that no other policies changed at the same time. To do this, we estimate specification \eqref{eq:1} using a sample of students two years before applying to high school. In the absence of other policies changing simultaneously, there should be no difference in the likelihood of changing addresses for these students. 
We find no response in the year prior to high school enrollment for this cohort as shown in Table \ref{tab:placebo.}. Our main result is thus robust to adding a third difference in a triple-difference design.

\begin{table}[h]
    \caption{Placebo regression results: The effect of high-school application incentives on moving behavior, two years before high school enrollment}
    \label{tab:placebo.}
     \begin{tabularx}{.95\linewidth}{Xc cc cc cccc}
    \toprule
     &\multirow{4}{*}{All}&&&\multicolumn{4}{c}{Parental income}\\
    &&&&\multicolumn{2}{c}{$\leq$p50}&\multicolumn{2}{c}{$>$p50}\\
    &&\multicolumn{2}{c}{Parental income}&\multicolumn{4}{c}{Middle school GPA}\\
    &&$\leq$p50&$>$p50&$\leq$p50&$>$p50&$\leq$p50&$>$p50\\
    &(1)&(2)&(3)&(4)&(5)&(6)&(7)\\
    \midrule
    Post $\times$ treated&       0.000&       0.001&      -0.001&       0.001&       0.001&      -0.001&      -0.001\\
                    &     (0.000)&     (0.001)&     (0.001)&     (0.001)&     (0.001)&     (0.001)&     (0.001)\\
MDV                 &       0.003&       0.004&       0.002&       0.004&       0.003&       0.003&       0.002\\
\midrule
Observations        &     163,514&      81,993&      81,453&      49,628&      31,716&      32,055&      48,795\\

    \bottomrule
    \end{tabularx}\\
      \begin{minipage}{0.95\linewidth}
   \footnotesize Notes: The table shows the coefficients from estimating equation \eqref{eq:1} using ordinary least squares. The dependent variable takes the value of 1 if the individual changed the address before March 1 in the  penultimate school year before high school enrollment, and 0 otherwise. Only individual address changes where no other member of the household changed address are considered. The variable treated is 1 for individuals living in municipalities where demand for high schools is in the top quartile (see Figure \ref{fig:geographvara} (a)) and zero otherwise. The variable post is one in all years after 2011. All regressions include the full set of year and municipality fixed effects, as well as the  full set of controls. The full set of controls includes a variable for the average parental years of schooling, a variable for the average parental disposable income, a variable for the 9th-grade GPA, and an indicator for the biological gender being female. We also include indicator variables for missing parental education, missing parental income, and missing 9th-grade GPA. Columns (2) to (7) show sub-sample regressions based on parental disposable income and middle school GPA, for both measures the median is calculated by cohort and municipality.         MDV is the mean of the dependent variable for untreated individuals.         Standard errors clustered at the municipality of residence at the beginning of the penultimate year before high school enrollment in parenthesis. 
    \end{minipage}
\end{table}

To assess whether we are capturing trends over time, Appendix Figure \ref{fig:trends} shows yearly averages in address changes between October 1 and March 1. The students residing in treated municipalities showed almost identical moving patterns to the students in the control municipalities prior to the reform in 2012, both in terms of levels, but also - and more importantly - in terms of trends. Since 2012 we observe a clear divergence between the two groups. Note that the increasing gap could be driven by two forces. First, a learning effect is where students learn about the new system over time. Second, a dynamic effect that reflects the possibility that if more students start to change their address, it increases the incentive for other students to change their address.

\begin{figure}[h!]
\begin{center}
    \includegraphics[width=1\linewidth]{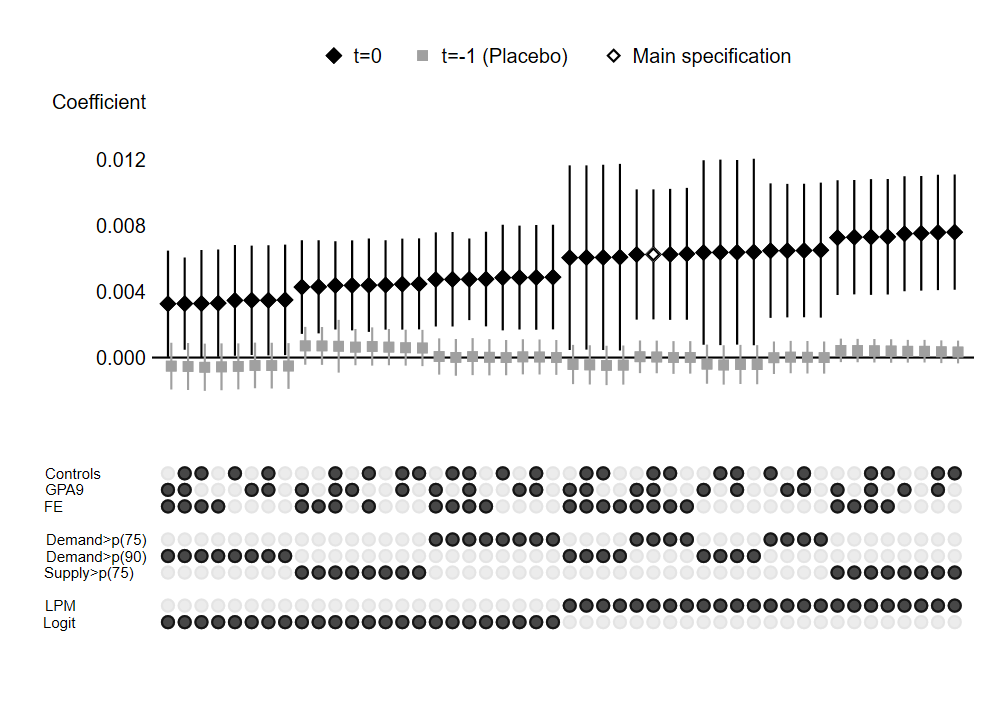}
\caption{Specification curve\vspace{12pt} \\ 
\begin{minipage}{1\linewidth}
\footnotesize
Notes: The black marker in the row for Controls indicate that the estimation includes the full set of controls. The black markers in row GPA9 indicate that controls for middle school GPA are included. The black markers in the row FE indicate that we include municipality and year-fixed effects (and grey markers that we only included pre and treatment indicators). The second panel shows various definitions of the treatment group. First, Demand$>$p(50) indicates that individuals are defined as treated if they reside in municipality above the median according to the demand measure described in Figure \ref{demand}.  Demand$>$p(75)  (the main specification) and Demand$>$p(90) uses the same measure, but respectively the 75th and 90th percentile. Supply$>$p(75) indicates that treatment is defined as the individual being in the top third of the distribution in terms of the number of high schools within 20Km. Finally, LPM and Logit indicate whether the specification is estimated with OLS as a linear probability model or as logit. 
\end{minipage}}
\label{fig:speccurve}
\end{center}
\end{figure}

Finally, we assess the robustness of our results to various alterations of the estimation specification. First, we consider the inclusion of control variables and municipality-fixed effects. Given our research design, we expect that including or not including these variables should not impact the point estimates. This is confirmed by Figure \ref{fig:speccurve}. As the first three rows below the chart show, there is no clear pattern showing that including more or fewer controls has a systematic effect on the point estimate. 

Moving to the next panel, we study the effect of alternative definitions of the treatment variable. First, we consider different thresholds for the municipality to be included in the treated group. We observe that the more municipalities are included in the treated group, the smaller the point estimate. This is not surprising, as broadening the definition of treatment will cause us to include more untreated in the treatment group and thus leading to a smaller estimated response. In the last row of the second panel, we show that using a supply-driven treatment definition is not systematically related to the size of the estimated coefficient. In this specification, we define treatment on an individual level based on the number of high schools within 20km.  Finally, given that we estimate probabilities close to zero, we show that the linear probability model provides similar results to a logit specification, but with slightly lower magnitudes.\footnote{In Appendix Table \ref{tab:res9} we also show that the conclusions are not changed when considering the full cohort finishing compulsory schooling.}

\subsubsection{Address manipulation \& application behavior}
\begin{figure}[h!]

\includegraphics[trim=1cm 1cm 1cm 1cm, clip,width=0.99\linewidth]{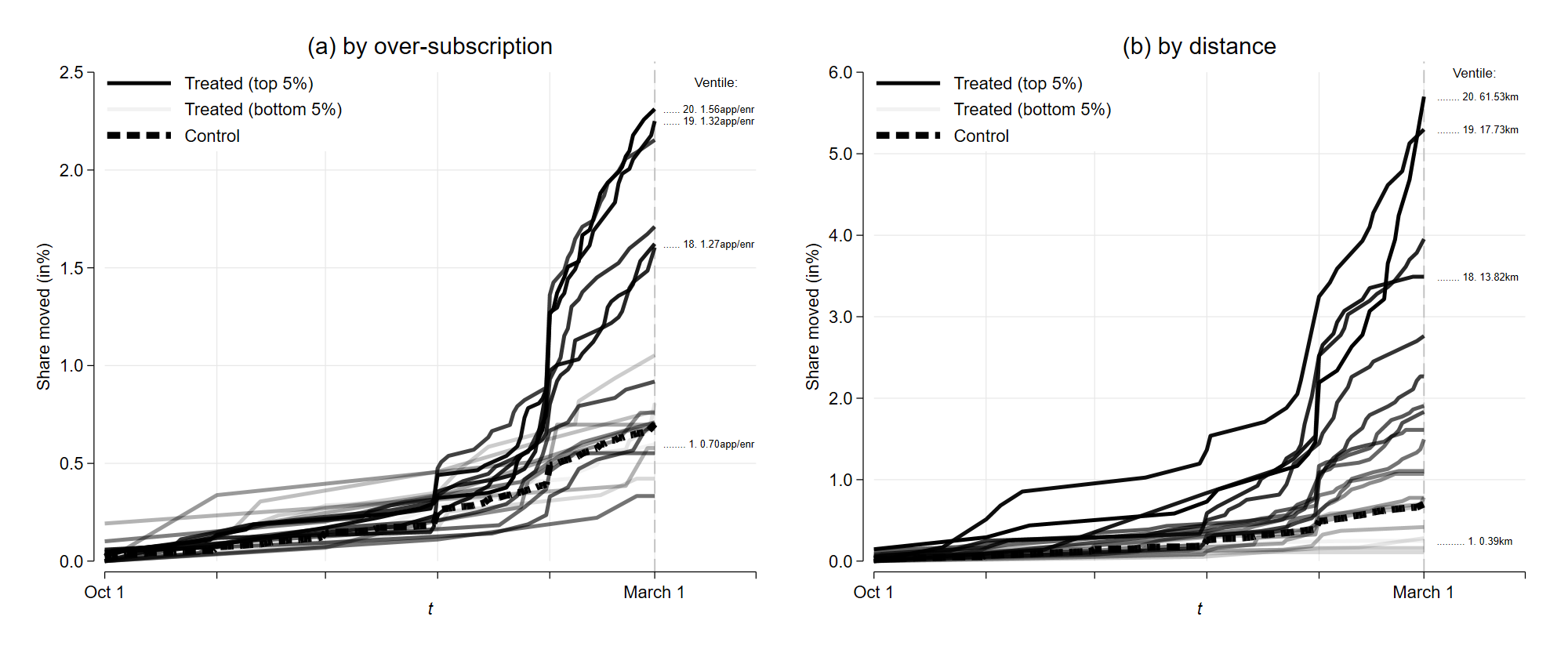}
\caption{Cumulative address changes by characteristics of the first listed high school\vspace{6pt} \\ 
\begin{minipage}{1\linewidth}
\footnotesize
Notes: The chart shows the cumulative address changes by respectively distance to the first listed high school and popularity (or how over-subscribed) of the first listed high school. The dashed lines show the control means, each of the solid lines are for  ventiles of the distance and popularity distribution. The lines are based on all students who applied to high school between 2012 and 2017 and only considers address changes where no other household member changed address.  
\end{minipage}}
\label{fig:behavior}
\end{figure}

Until now we exploited variation in when the applicants finished compulsory schooling and their initial address to identify the response to such incentives. However, this incentive is only there if the individual actually applies to an oversubscribed high school and if the (potential) new address is closer to the oversubscribed high school. We refrained from using this information for the main identification exercise because both these factors depend on the high school the applicant applies to, and it is therefore endogenous to the ability to change or not change address. In Figure \ref{fig:behavior} we show that the increase in address changes is considerably larger for the individuals where we expect to see larger responses. Up to 5 percent of the moves, which is more than double the average response shown in Appendix Figure \ref{fig:ab}. We also see much stronger responses from students who apply to the most popular high schools. Moreover, students who reside in treated areas, but apply to schools that did not receive more applicants than capacity did not show different behavior than the students residing in the control municipalities. 

The regression results in Table \ref{tab:behavior} confirm the graphical evidence from Figure \ref{fig:behavior}. We observe strong responses by the top 40 percent in terms of distance to the school.  Moving to panel B we observe significant responses by applicants to the 60 percent most popular high schools.

\begin{table}[h!]
    \caption{Regression results by distance to first priority and popularity of the first priority high school}
    \label{tab:behavior}
    \begin{tabularx}{.95\linewidth}{cXc cc cc ccc}
    \toprule
    &&\multicolumn{6}{c}{Position in the distance and popularity distribution}\\
   &&1-20\%&21-40\%&41-60\%&61-80\%&81-100\%&96-100\%\\
    &&(1)&(2)&(3)&(4)&(5)&(6)\\
    \midrule
    \multicolumn{7}{l}{\emph{A. By distance to school listed first}}\\
    &Post $\times$ treated&      -0.002&       0.002&       0.002&       0.010&       0.019&       0.018\\
&                    &     (0.001)&     (0.002)&     (0.003)&     (0.005)&     (0.004)&     (0.009)\\
&MDV                 &       0.002&       0.005&       0.009&       0.013&       0.026&       0.041\\
&Distance (Km)                &       0.876&       1.996&       3.195&       5.129&      14.906&      32.818\\
&Observations                   &     104,195&     104,195&     104,197&     104,188&     104,188&      95,991\\
\\
    \multicolumn{7}{l}{\emph{A. By popularity of school listed first}}\\
   &Post $\times$ treated&      -0.000&       0.002&       0.014&       0.012&       0.014&       0.008\\
 &                   &     (0.001)&     (0.002)&     (0.006)&     (0.003)&     (0.003)&     (0.007)\\
&MDV                 &       0.006&       0.006&       0.012&       0.014&       0.017&       0.009\\
&App/Enr              &       0.789&       1.014&       1.125&       1.224&       1.669&       3.025\\
&Observations                   &     102,977&     103,164&     102,673&     102,907&     101,745&      92,719\\\\[-12pt]
    \bottomrule
    \end{tabularx}\\
      \begin{minipage}{0.95\linewidth}
        \footnotesize Notes: The table shows the coefficients from estimating equation \eqref{eq:1} using ordinary least squares. The dependent variable takes the value of 1 if the individual changed the address before March 1 in the  school year before high school enrollment, and 0 otherwise. Only individual address changes where no other member of the household changed address are considered. The variable treated is 1 for individuals living in municipalities where demand for high schools is in the top quartile (see Figure \ref{fig:geographvara} (a)) and zero otherwise. The variable post is one in all years after 2011. All regressions include the full set of year and municipality fixed effects, as well as the  full set of controls. The full set of controls include a variable for the average parental years of schooling, a variable for the average parental disposable income, a variable for the 9th-grade GPA, and an indicator for the biological gender being female. We also include indicator variables for missing parental education, missing parental income, and missing 9th-grade GPA.   MDV shows the mean of the dependent variable for untreated individuals.        Standard errors clustered at the municipality of residence at the beginning of the last year before high school enrollment in parenthesis.
    \end{minipage}
\end{table}

Before we turn to the implications of address manipulation it is useful briefly to consider how the students are able to change their addresses. First, in Appendix Figure \ref{fig:addressmanipulationfull} and the corresponding Appendix Table \ref{tab:extensive} we assess whether the identified response is driven by students changing address earlier than they otherwise would. It could be the case that both pre- and post-students in the treated areas are more likely to change address before high school start to reduce commuting time or other motives. After the reform, there is an incentive to announce this move earlier to be aligned with the admission deadline. In that case, the response we have identified would be driven by the timing of address changes. We study this by looking at all address changes up until high school actually starts. As we show in the appendix this is not the case. All of the overall response is driven by address changes that would otherwise not have happened. 

In Appendix Table \ref{tab:relative} we decompose the response into whether the address change was to an extended family member. Anecdotally using relatives' addresses is a  common strategy for address manipulation \citep{guard} We therefore decomposed the response into address changes to an address where a relative of the focal applicant already lived and all other address changes. Indeed, as the table shows, about half of the response is to relatives. However, this does not shed light on whether the address change reflects a real move or not. To assess this we exploit the 2018 reform in the next subsection.

\subsubsection{Reducing access to address manipulation}
So far we have documented that the 2012 reform increased the incentive to manipulate ($w$ in our theoretical model.) In line with theory, the reform led to an increase in address changes driven by individual moves. We now turn to the second reform described in section \ref{sec:inst}, where the municipality of Copenhagen in 2018 by default rejected individual address changes by minors. This reform, therefore, decreased the weight ($w$) on the manipulable part in the eligibility score and increased the cost of deceiving ($\gamma$) by imposing a hassle cost on relocated applicants\footnote{The hassle costs may come from applicants trying to document the realness of their move or from applicants actually moving when they may not have done so before the reform}, but only for applicants who wished to manipulate their score by changing the address to a location in Copenhagen. One challenge in studying this reform is that the treatment is defined based on the outcome variable as individuals are treated who wanted to change their address to an address in the municipality of Copenhagen. We, therefore, defined everyone as treated by the 2018 reform who lived around or in Copenhagen at the beginning of the last school year. These are students who would be likely to apply to schools in the municipality of Copenhagen, but we do not condition on applying.  

\begin{figure}[h!]
\includegraphics[width=\linewidth]{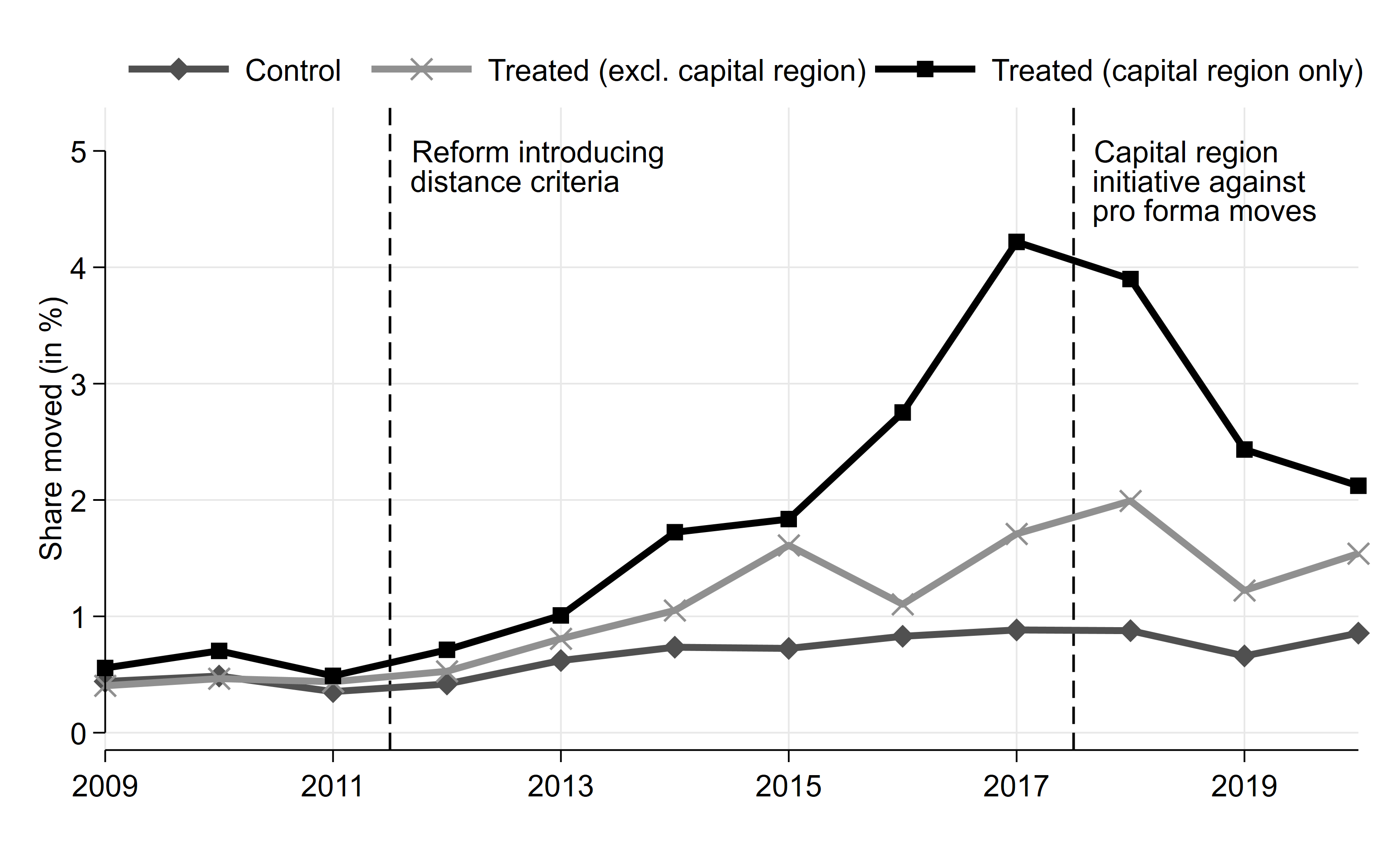}
\caption{Address changes between October 1 and March 1 over time\vspace{6pt} \\ 
\begin{minipage}{1\linewidth}
\footnotesize
Notes: The figure shows the share of students who changed their address between October 1 and March 1 in the last year before high school enrollment for the treated and control groups. Only address changes where no other member of the household change the address are considered. 
\end{minipage}}
\label{fig:trendsl4}
\end{figure}

In Figure \ref{fig:trendsl4}, we show the development in address changes until 2020, where we filtered out students initially residing in the greater Copenhagen area. As expected we observe a strong increase in address changes by individuals living in the Copenhagen (capital) area until 2017 in line with the results studied so far. However, we also see a clear drop in address changes around Copenhagen after 2017. The decline in address changes after 2017 suggests a) that the reform was effective and b) that address changes were mainly proforma in the sense that checks for whether the address change reflects a real move reduced the number of address changes.\footnote{In Appendix Figure \ref{fig:trendsl4_appendix} we show a chart of address changes in a municipality relative to the initial population, where we define the treated group as only the municipality of Copenhagen. The patterns of that figure are very much in line with the patterns in Figure \ref{fig:trendsl4}. } 

To estimate the effect of the 2018 reform, we use a similar approach as above and estimate equation \ref{eq:1}, with the following two modifications. First, the pre-period is defined as 2017 and the post-period is the period from 2018 to 2020.\footnote{See Appendix Table \ref{tab:sumstat2018} for summary statistics for this group.} Second, the treated group is defined as students residing in the greater Copenhagen area, and the control group is the same as the control group above. Students residing in treated municipalities outside Copenhagen are not included in this analysis.  The results are shown in Table \ref{tab:2018reform} and confirm that individual moves declined significantly after 2017 and that the response is mostly driven by students from high-income families. The magnitudes of the response to the 2018 reform are in line with canceling out the effects of the initial response of the 2012 reform. 

\begin{table}[h!]
    \caption{Regression results: The effect of the 2018 reform for reducing address manipulation}
    \label{tab:2018reform}
     \resizebox{\linewidth}{!}{\begin{tabularx}{1.05\linewidth}{cXc cc cc ccc}
    \toprule
   &&\multirow{4}{*}{All}&&&\multicolumn{4}{c}{Parental income}\\
    &&&&&\multicolumn{2}{c}{$\leq$p50}&\multicolumn{2}{c}{$>$p50}\\
    &&&\multicolumn{2}{c}{Parental income}&\multicolumn{4}{c}{Middle school GPA}\\
    &&&$\leq$p50&$>$p50&$\leq$p50&$>$p50&$\leq$p50&$>$p50\\
    &&(1)&(2)&(3)&(4)&(5)&(6)&(7)\\
    \midrule
    \multicolumn{7}{l}{\emph{A. Individual address changes}}\\
    &Post $\times$ treated&      -0.008&      -0.005&      -0.012&      -0.003&      -0.008&      -0.006&      -0.016\\
&                    &     (0.003)&     (0.004)&     (0.005)&     (0.003)&     (0.007)&     (0.005)&     (0.007)\\
&MDV                 &       0.007&       0.007&       0.007&       0.007&       0.008&       0.007&       0.007\\
\\
    \multicolumn{7}{l}{\emph{B. Household address changes only}}\\
    &Post $\times$ treated&      -0.000&       0.003&      -0.004&       0.001&       0.006&      -0.003&      -0.004\\
&                    &     (0.003)&     (0.004)&     (0.004)&     (0.006)&     (0.007)&     (0.007)&     (0.005)\\
&MDV                 &       0.029&       0.032&       0.025&       0.034&       0.029&       0.031&       0.021\\
\midrule
\multicolumn{2}{l}{Observations}        &      52,124&      26,136&      25,960&      16,155&       9,882&      10,001&      15,843\\\\[-12pt]
    \bottomrule
    \end{tabularx}}\\
      \begin{minipage}{1\linewidth}
      \footnotesize Notes: The table shows the coefficients from estimating equation \eqref{eq:1} using ordinary least squares. The dependent variable takes the value of 1 if the individual changed the address before March 1 in the   school year before high school enrollment, and 0 otherwise. Only individual address changes where no other member of the household changed address are considered. The variable treated is 1 for individuals living in municipalities where demand for high schools is in the top quartile (see Figure \ref{fig:geographvara} (a)) and zero otherwise. The variable post is one in all years after 2011. All regressions include the full set of year and municipality fixed effects, as well as the  full set of controls. The full set of controls include a variable for the average parental years of schooling, a variable for the average parental disposable income, a variable for the 9th-grade GPA, and an indicator for the biological gender being female. We also include indicator variables for missing parental education, missing parental income, and missing 9th-grade GPA.  Panel A. shows results where the dependent variable is 1 if the individual student changed address in the period between October 1 and the last day of February in the school year before high school enrollment, and 0 otherwise. In Panel B only address changes where no other member of the initial household changes address are considered. In Panel B only address changes where at least one other member of the initial household changed address are considered. Columns (2) to (7) show sub-sample regressions based on parental disposable income and middle school GPA, for both measures the median is calculated by cohort and municipality.         MDV shows the mean of the dependent variable for untreated individuals.         Standard errors clustered at the municipality of residence at the beginning of the last year before high school enrollment in parenthesis. 
    \end{minipage}
\end{table}

The pattern shown in Figure \ref{fig:trendsl4} and the regression results from Table \ref{tab:2018reform} give us two insights. First, that a simple policy can reduce manipulation, and second, that the initial increase in the manipulation was driven by proforma moves and not real moves because otherwise, we would not have seen a comparable reduction in the response. 

A note on the cost of policy enforcement is appropriate. While we do not have a monetary measure of the cost, we performed an interview with a caseworker responsible for the processing of moving requests. The change meant that a considerable number of moves had to be addressed on an individual basis and the administrative unit was put under considerable time pressure. It is an open question to which extent the benefits of eliminating manipulation outweigh the cost of enforcement.

\subsubsection{Implications for justified envy}
We have so far shown that students act according to our theoretical expectations and move to a larger degree when given the incentive. For a fixed supply of seats in oversubscribed schools, this manipulation may crowd out other applicants. We, therefore, now turn to the implications for access to schools for all applicants. 

\begin{figure}[h!]
\includegraphics[width=1\linewidth]{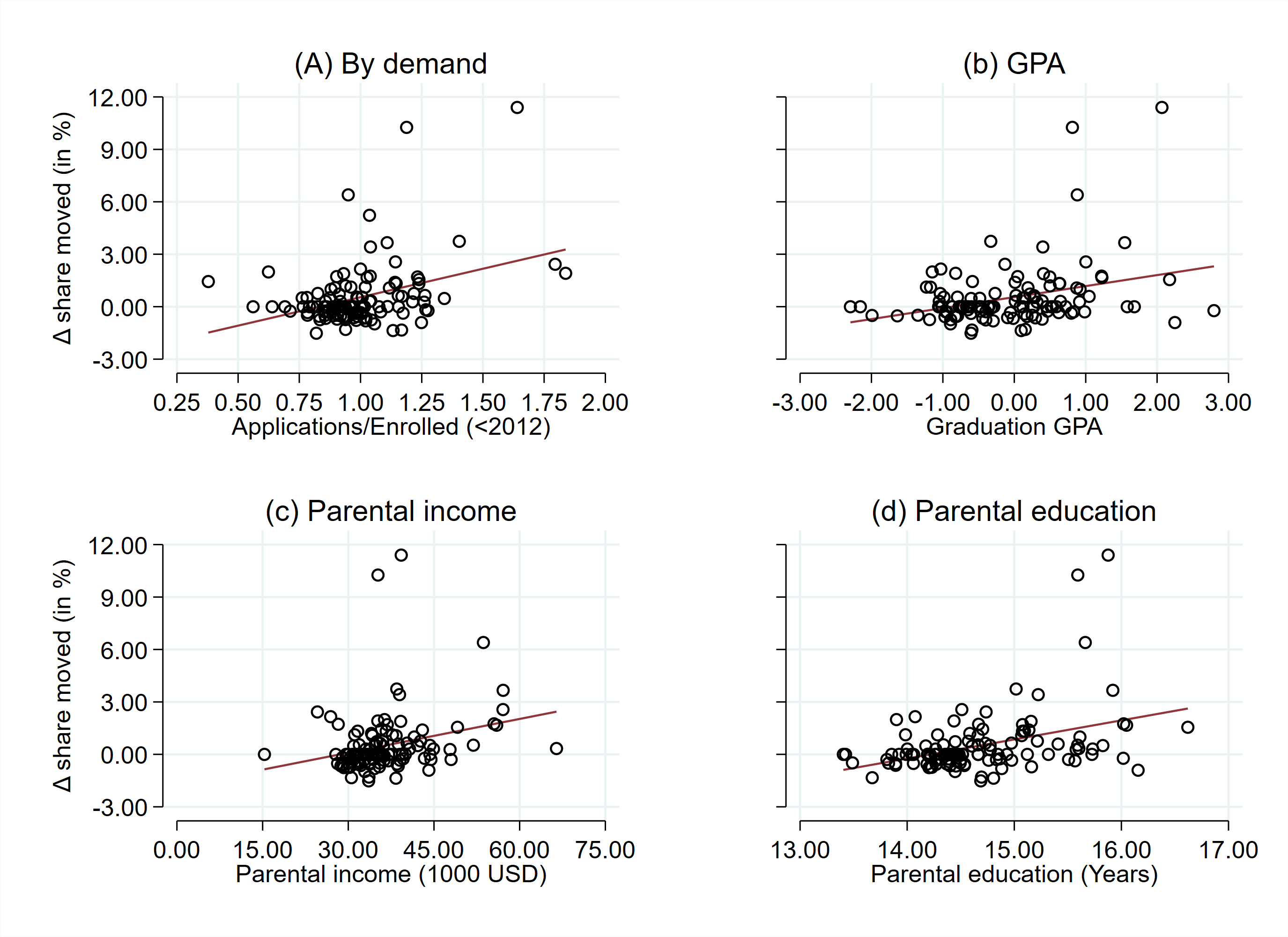}
\caption{High school level differences in address changes, difference between the level in 2017 minus the average for the years 2009 to 2011 \vspace{6pt} \\ 
\begin{minipage}{1\linewidth}
\footnotesize
Notes: Each circle shows the average for a high school. Graduation GPA, Parental Income, and Parental Education are measured for the graduates in the year before the share of moves is measured. 
\end{minipage}}
\label{fig:howcommondif}
\end{figure}

As Figure \ref{fig:behavior} and Table \ref{tab:behavior} clearly showed that the effects are driven by the most popular high schools, the behavior will likely affect access to these schools. Indeed, in Figure \ref{fig:howcommondif} we show the high school level average change in individual address changes up to the high school application deadline. The Figure reveals that the change in individual address changes in correlated with how over-subscribed the high school is (a), the high school graduation GPA (b), the average parental income (c), and the parental years of education (d). Moreover, the figure shows that at some high schools, the change in moving share is larger than ten percentage points. Given that the baseline rates were close to zero, the changes are close to the levels, the values  suggest that in some high schools, more than one out of ten students strategically changed their address prior to the application deadline (which is also confirmed by Appendix Figure \ref{fig:howcommonlevels} showing the levels).

We therefore now assess whether address manipulation also leads to justified envy as predicted by our theoretical result in section \ref{sec:theory}. Empirically we define justified envy as follows. Firstly, the student did not enroll in the school listed first in their rank order list and the school received more first-priority applicants than they enrolled. Secondly, the student lived closer to the school than the initial address of an admitted student who changed their address, but further away than the new address of the admitted student. 

A challenge in this exercise is that not all address changes are manipulation. Our goal is only to identify justified envy caused by address changes in reaction to the application incentives.  We identify the probability that a move is strategic by invoking a monotonicity assumption: We assume that the probability of moving does not decrease when the incentive is introduced.  Under the monotonicity assumption, $\beta_3$ in equation \eqref{eq:1} ( the coefficient on the difference-in-differences estimate) reflects the share of applicants in the treated region which move when given the incentive, $Pr(manipulation|post_i=1,treat_i=1)$. Following \cite{abadie2003semiparametric} we identify the share of moves which are manipulation for the treated regions:
\begin{align}
    Pr(manipulation|move_i=1) =\frac{\beta_3}{E[moved|post_t=1,treated_i=1]}.\label{eq:share_move_str}
\end{align}

To capture heterogeneity in manipulation behavior, we consider the separate coefficients from eight different regressions using all combinations of female, parental income above the municipality median and parental education above the median (see Appendix Table \ref{tab:simulationreg}). For sons of above median income and above median education parents, the difference-in-differences coefficient is 0.013 and given that 19 percent of this group changed address in the post period, equation \ref{eq:share_move_str} implies that 69.6 percent of the address changes in this group were manipulative. For boys in the below-median income and education groups, the corresponding share is only 6.1 percent.   We initiate a simulation by assigning a draw from a standard uniform distribution to each address change. If the draw is below the estimated probability, the move is set to be manipulated. If the draw is above the estimated probability, we let the applicant stay at the new address. After having done this for all address changes we calculate the share with justified envy because of manipulative address changes.  We repeat this procedure 150 times.

\begin{figure}[h!]
\subfloat[Distribution across 150 repetitions \label{envya}]{\includegraphics[width=0.49\linewidth]{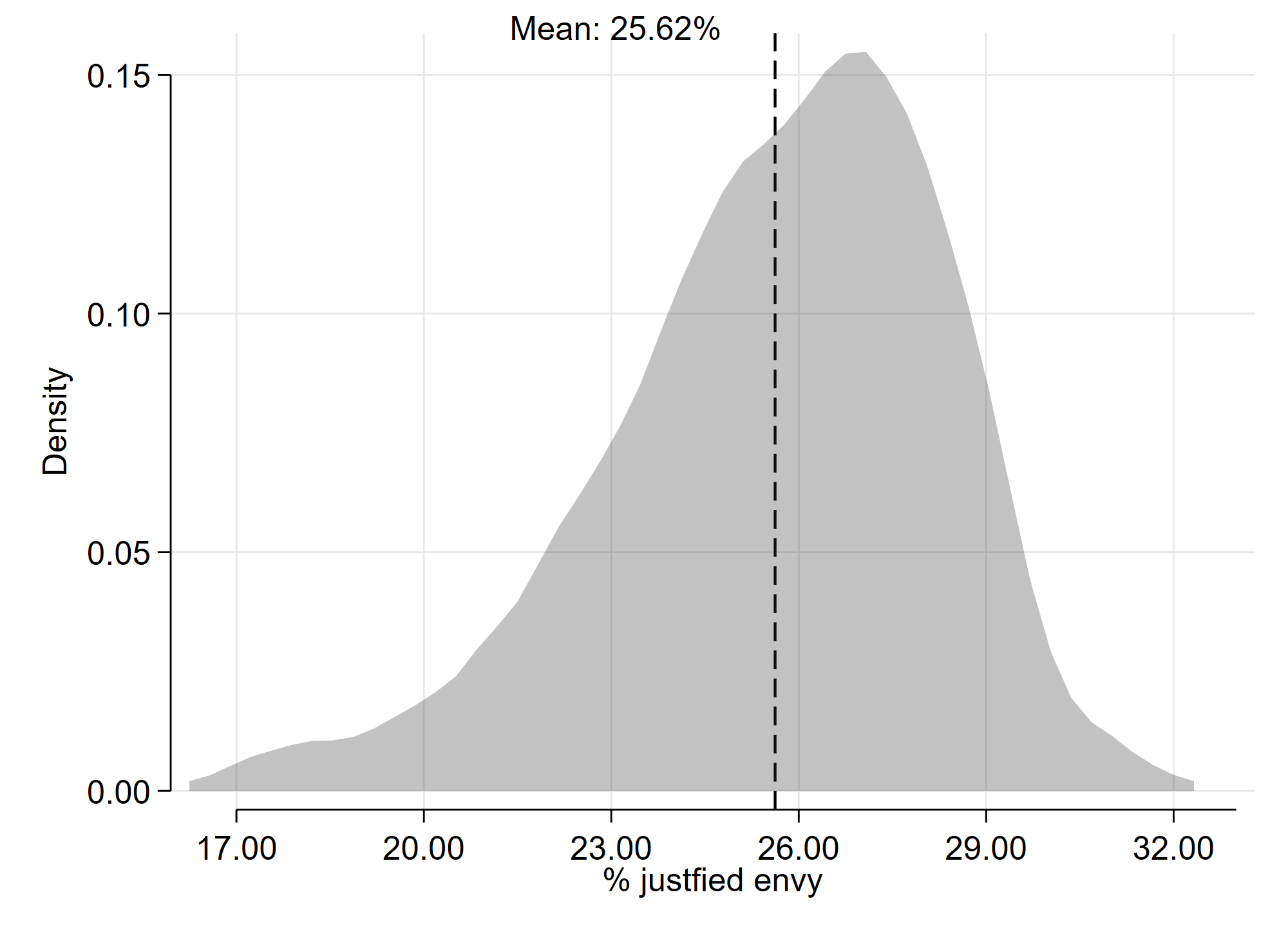}}
\subfloat[Average justified envy by high school\label{envyb}]{\includegraphics[width=0.49\linewidth]{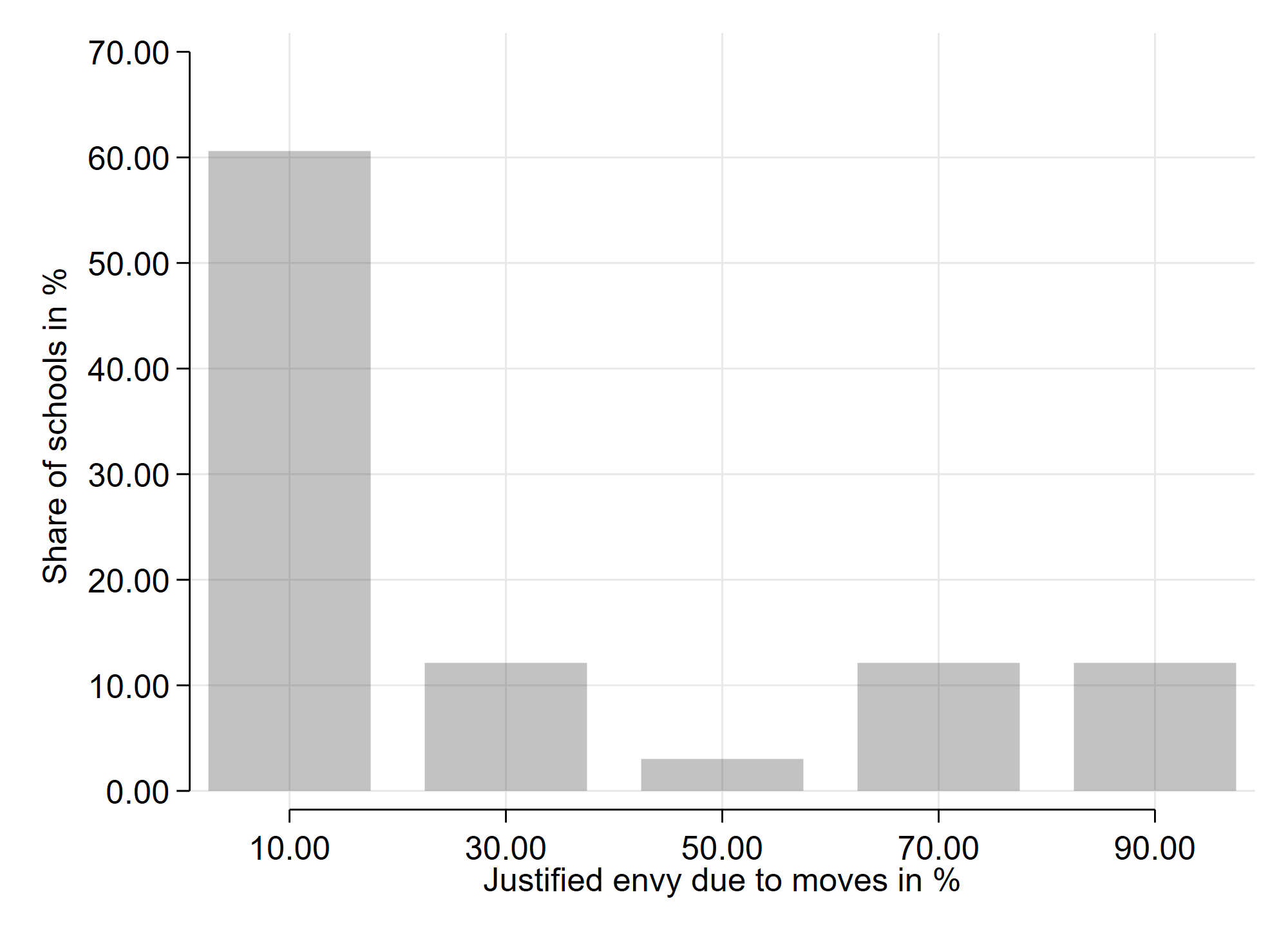}}
\caption{Justified envy because of address manipulation\vspace{6pt} \\ 
\begin{minipage}{1\linewidth}
\footnotesize
Notes: Figure \ref{envya} shows the distribution of the share of applicants with justified envy conditional on not enrolling, across the 150 iterations.  Figure \ref{envyb} shows the average share with justified by high school across 150 iterations. The shares are calculated for the year 2017.
\end{minipage}}
\label{fig:envy}
\end{figure}

Figure \ref{envya} shows the distribution of justified envy across 150 simulation iterations. The share of non-enrolled applicants with justified envy ranges between 18 and 33 percent with an average of 26 percent. In other words, on average more than one out of four students not enrolling in their first listed high school have justified envy towards a student who manipulated their eligibility by changing address. And the fact that all iterations were between about one-fifth and one-third suggest that the uncertainty around this number is limited. However, this means conceals substantial heterogeneity across schools as highlighted by Figure \ref{envyb}. For some schools, more than 90 percent of applicants who did not enroll have justified envy because of address manipulation. Consequently, address manipulation might affect the peer group of rejected applicants. We study this in the next section.

\subsubsection{Simulation of alternative allocations}
We have shown that 26 percent of rejected applicants have justified envy. However, multiple rejected applicants may have justified envy towards the same admitted applicants. Thus, from this number, we cannot directly deduce the distributive consequences of manipulation.
To assess the consequences of manipulation we need to compare the equilibrium allocation to a counterfactual allocation, where manipulation is limited.
We, therefore, construct a simplified simulation model of the secondary school admission system.\footnote{An alternative to simulation is to formulate our difference-in-difference model as a first stage and estimate corresponding reduce forms models. However, because capacity is limited, applicant strategies affect other applicants, which breaks the Stable Unit Treatment Value Assumption. This implies that one cannot use IV-methods to estimate the consequences of strategic moves.} 

Concretely, we repeat the approach from the last subsection to classify moves as manipulation. However, we now simulate the effect of removing these moves, just like the 2018 reform in Copenhagen did. In practice, we set the address of applicants classified as manipulating to their original address and we then reallocate all students given the schools' applicants and the rank order lists using Immediate Acceptance.  We assume that the policy comes as a surprise to applicants. We do so because choosing a location and submitting preferences are simultaneous decisions. In addition, the mechanism is not strategy-proof which means that modeling this joint decision would require complicated and restricted structural modeling. We circumvent this modeling by letting the authorities shut down moves after preferences have been submitted. However, this introduces changes in expectations for the applicants next year, which we ignore.

For each individual, we compute the probability of going up or down the rank order list across the simulations. This allows us to compute the average number of winners (getting a higher priority) and losers (getting a lower priority). We also characterize winners and losers by weighting covariates by probabilities of being a loser or winner respectively. We present the results of this exercise in Table \ref{tab:sim_WL_indi}. Unsurprisingly the winners of removing manipulative moves are children of parents with lower parental income than the losers. In the baseline scenario, the winners have high school peers with prior attainment of 0.43SD. After the reallocation, the winners have peers with a prior attainment that is 0.19SD higher. Address manipulation, therefore, pushes applicants out of desired schools into schools with less able peers. 

\begin{table}[ht]
        \begin{tabularx}{0.75\linewidth}{cXcccc}
                \toprule
                {} && Unaffected &  Losers& Winners \\
                \midrule
            &Female&  0.61&  0.60&  0.55\\
&Parental income (1,000 USD)& 52.08& 62.57& 58.28\\
&Peer GPA (baseline)&  0.60&  0.67&  0.43\\
&Peer GPA (counter-factual)&  0.60&  0.52&  0.62\\
&Applicants/Enrolled&  1.04&  1.38&  1.34\\
\vspace{-12pt}\\
            \bottomrule 
                \end{tabularx}
    \caption{Characteristics of applicants based on simulation results}\label{tab:sim_WL_indi}

      \begin{minipage}{0.75 \linewidth}
        \footnotesize Notes: The table shows the simulation results based on 150 simulations where in each iteration we first classify whether a given address changes as manipulation and then move manipulative movers back to their original address. We then reallocate all students to high schools given their rank order list and high school capacity. The table shows the share of students accepted at a higher-ranked school than in the baseline (winners), the students accepted at a lower-ranked school (losers), and those unaffected by removing all manipulative moves. 
    \end{minipage}
\end{table}

\subsubsection{Survey evidence}

To shed light on the applicants' considerations and level of information we conducted a survey for the 2020 and 2021 high school applicants. Given the empirical findings above we would expect applicants in treated areas to be better informed about the mechanism and the survey results thus serving as a sanity check of the patterns we observe in the administrative records.\footnote{The survey was conducted after the 2018 address manipulation reform in Copenhagen which limited the scope for manipulation in the metropolitan area.}

In both surveys, around 4 percent of the respondents answered that they had changed their address during the three months preceding the application deadline of March 1st. Furthermore, approximately 13 percent answered that they had considered changing their home address to increase their chances of getting admitted to their first-ranked high school.

Of those in the 2021 survey, who answered that they considered changing their address, we asked what spoke against and in favor of doing so.\footnote{The respondents are faced with a battery of answer categories of which they can choose more than one answer, see question Q3 and Q4 in Appendix Table \ref{app:survey}} Of reasons not to move, the majority answered that they felt sufficiently certain that they would get admitted at their first priority and hence did not see the need to move anyway (32 percent). The next most popular answers were that they were afraid of getting caught by the authorities (22 percent) and/or did not have a place to move to (16 percent). The least important were the opinions of family members, friends (6 percent), and prospective classmates (7 percent). 

Of the reasons to move, the most popular answer was that the respondent felt like he/she would fit in better at the particular high school (52 percent). Otherwise, the characteristics of the first prioritized high school seem to be more appealing in terms of supplied courses (43 percent), academic level (33 percent), and/or social activities (31 percent). Lastly, respondents valued the area close to the high school (14 percent), the fact that friends secured a seat at the particular high school (16 percent) or the opportunity to reduce commuting time to the high school (33 percent).    

\begin{figure}[ht!]
\subfloat[Eligibility Score\label{demand}]{\includegraphics[width=0.49\linewidth]{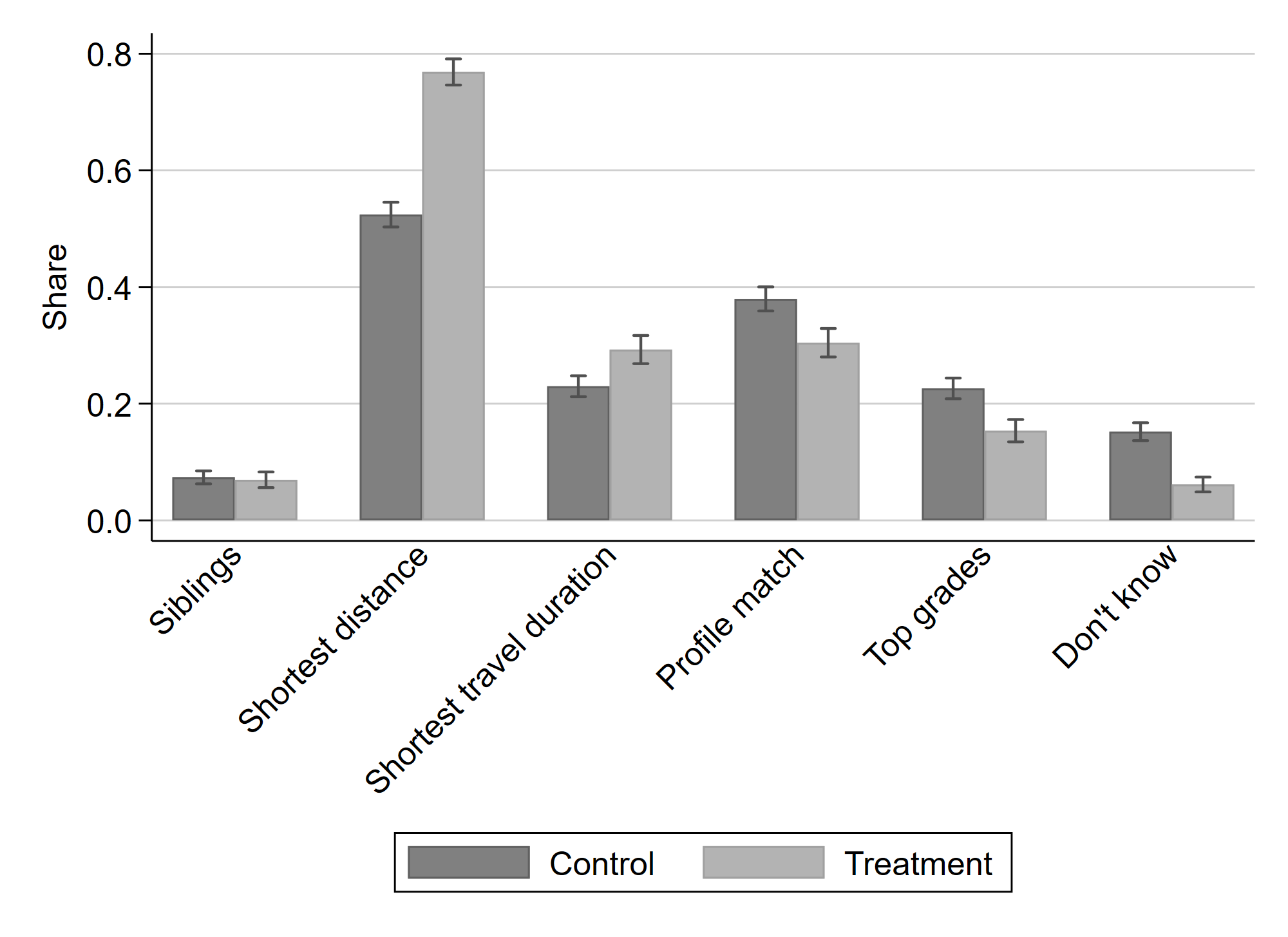}}\hfill
\subfloat[Immediate Acceptance Mechanism\label{supply}]{\includegraphics[width=0.49\linewidth]{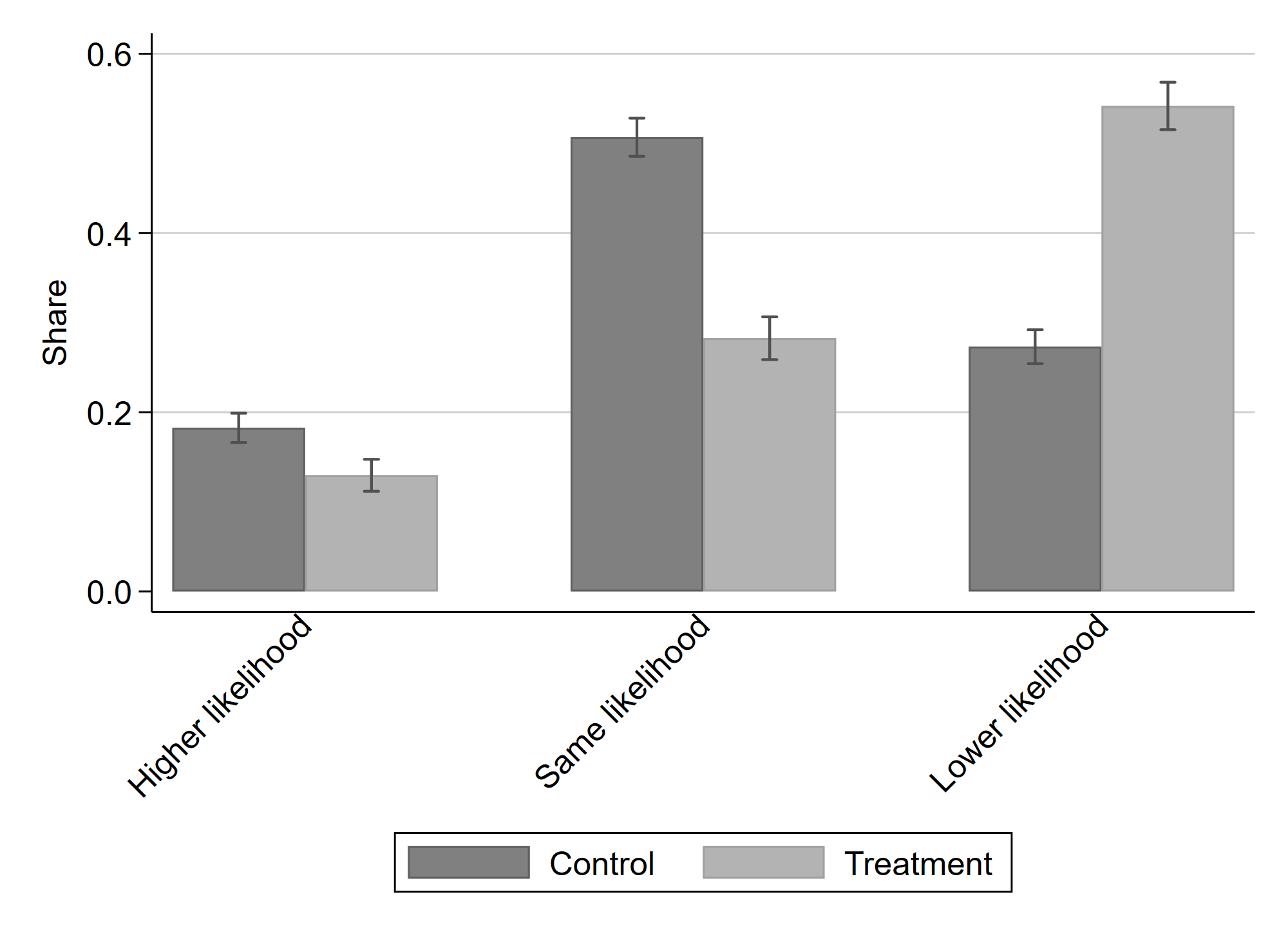}}
\caption{Knowledge of the Eligibility Score and Understanding of Immediate Acceptance Mechanism \vspace{6pt} \\ 
\begin{minipage}{1\linewidth}
\footnotesize
Notes: The figure shows the distribution of answers across answer categories for respondents in the treatment and control groups in the year 2020.  In panel (a) we ask about what determines eligibility. In panel (b) we ask them whether listing a school as first priority increases the likelihood of admission. The question wordings can be found in table \ref{app:survey} as Q6 and Q5, respectively.  
\end{minipage}}
\label{fig:mech1mech2}
\end{figure}

In Figure \ref{fig:mech1mech2}, we check whether the respondents' knowledge and understanding of the mechanism depends on whether they belong to the treatment or control group.\footnote{Treated respondents are those residing in a geographic area in the top 25 percent of the over-subscription distribution as in the main analysis} In both cases, we find that treated respondents answer correctly more often than the control respondents. As information about the mechanism is not as crucial for the control group as it is for the treatment group, it corroborates the interpretation of our empirical findings as applicants react to the incentives of the mechanism.  

In the surveys, we asked respondents a question regarding their knowledge of the eligibility score and a question about their understanding of the immediate acceptance mechanism. In the 2020 survey, approximately 66 percent answered correctly, when asked how schools prioritize between applicants if there are more applicants than available seats. Furthermore, when we asked them whether they would have the same, lower or higher likelihood of getting admitted at their second priority in case of rejection at their first priority, only 33 percent answered that the likelihood would be smaller.

Together these survey results suggest that students in the treatment areas are better informed about how the assignment mechanism works, including that distance to the school is the main ingredient in the eligibility score. Moreover, the results show that the risk of getting caught is the main reason for not manipulating the address.   
\section{Conclusion}
\label{sec:conclusion}
In this paper, we have shown that incentives for manipulating moves work and that this has distributive consequences for allocations to school.

We developed a theoretical framework that describes applicant manipulation of school choice mechanisms when schools prioritize between applicants based on an eligibility score. We showed that applicants have an incentive to manipulate the priorities of schools if the schools are oversubscribed and the eligibility score depends on some manipulable component. This type of applicant deception differs from applicants manipulating their stated preferences since it leads to justified envy irrespectively of whether the school choice mechanism is the immediate acceptance or the deferred acceptance mechanism.

Using insights from the theoretical framework, we investigated whether applicants to Danish high schools reacted to a change in the incentive to deceive induced by a policy reform in 2012. We used a causal research design based on the difference-and-difference approach and found that the prevalence of applicants moving right before the application deadline increased by around 100 percent for those initially residing in geographic regions with a high share of oversubscribed high schools. Our results survive numerous robustness checks. Additionally, we found that this behavior was most prevalent among high schools where students graduated with high GPAs and among children of highly educated parents. We found that such behavior crowds out applicants from less privileged backgrounds, and therefore implies regressive redistribution.

Our findings suggest that school choice research must broaden its scope beyond the mechanism. Strategic actions go beyond stating preferences and manipulation of eligibility has real consequences for access to education.

\bibliography{references}

\begin{appendix}
    \numberwithin{table}{section}
    \numberwithin{figure}{section}
    \setcounter{figure}{0}
    \setcounter{table}{0}
    \setcounter{section}{0}
    \setcounter{subsection}{0}
    \renewcommand{\thetable}{A.\arabic{table}}
    \renewcommand{\thefigure}{A.\arabic{figure}}
    \renewcommand{\setthesection}{\Alph{section}}
    \renewcommand{\setthesubsection}{\Alph{subsection}}
    \renewcommand{\thetable}{A.\arabic{table}}
    \renewcommand{\thefigure}{A.\arabic{figure}}
    \appendix
    \clearpage
    \FloatBarrier
    \renewcommand{\theequation}{A.\arabic{equation}}
\section{Auxiliary theoretical analysis and concepts}\label{app:theory}

\subsection{Proofs of strategic properties}\label{app:proofs_strategic}

In the two proofs below we use the fact that after the students have decided on their preparatory actions then the priorities over students can be considered exogenous. As a consequence, the framework corresponds exactly to a standard school choice problem.

\noindent\textbf{Proof of Proposition \ref{proposition:manipulability}}: 
Pick any combination of actions by the students. As our framework is identical to standard school choice when student have taken their actions, it follows directly that DA is strategy-proof \citep{dubins1981machiavelli,roth1982economics} while IA is not \citep{abdulkadirouglu2003school}. 
Example \ref{example:strategic} provides a situation where student 1 uses deception for its own utility gain and thus deception-proofness is violated.
$\blacksquare$
\bigskip

\noindent\textbf{Proof of Proposition \ref{proposition:envy_free}}: 
Again, pick any combination of actions by the students. As our framework is identical to standard school choice when student have taken their actions, it follows that elimination of justified envy holds for DA \citep{abdulkadirouglu2003school} as it evaluates envy using priority conditional on preparatory actions.

To verify that invariant justified envy property can be violated, see Example \ref{example:strategic}. In the example, student 2 has justified envy when we use priority levels under the assumption that $w=0$ and thus invariant justified envy does not hold.
$\blacksquare$
\bigskip

The proof of Proposition~\ref{claim:monotone_deception} is found in the following sup-appendix as it requires an extended conceptual framework.

\subsection{Supply and demand framework}\label{app:supply_demand}

In this appendix, we introduce auxiliary market-like concepts by extending the demand and supply framework in  \citet{azevedo2016supply} to our setting. This allows us to construct individual demand functions for a given cutoff and demonstrate its properties. This is possible by simultaneously characterizing the optimal deception / preparatory action as a function of school cutoffs. 
The cutoffs have the role of prices as they determine what schools students can gain entry at. 

We begin extending the framework from the main text. Here, we define students in terms of a type $\theta\in\Theta$. 
We assume that each student type has exactly $K$ actions, including the null action (i.e., $|A_\theta|=K$). 
To allow for deception we assume $K\ge2$.
The set of types is defined as $\Theta=\mathbb{R}^|\mathcal{S}|\times [0,1]^{|\mathcal{S}|\cdot K}$.

To construct measures of demand function in our setup we require a way of measuring feasible choices. To do this, we construct an auxiliary measure that captures the action that maximizes admission entry at school $s$ for type $\theta$ if it deceives. 
The entry maximizing action, $a_{\theta,s}^{*}={\arg\min}_{a_\theta\in A_\theta}P_{\theta,s}(a_\theta)$ is the one that obtains the minimal priority score at school $s$. We say that student $\theta$ can afford school $s$ under cutoffs $P$ if the school is willing to admit $\theta$: affording $s$ without deception requires that $p_{\theta,s}(a_{\theta}^{0})\ge P_s$ and with deception that $p_{\theta,s}(a_{\theta,s}^{*})\ge P_s$. 

\subsubsection{Formal definition of demand and optimal deception}\label{app:demand_formal}

A student’s \textbf{demand} given cutoffs is her favorite school in terms of net utility among those she can afford. In other words, if students of type $\theta$ demand a school $s$ given cutoffs $P$ then they expect admission there. If no schools are affordable, define $D^\theta(P,w)=\emptyset$, meaning that the student demands are unmatched. We say that demand occurs with  \textbf{deception} if admission requires deception. We note that the individual demand does in fact capture the best response in terms of submitted preferences and preparatory action for every agent given cutoffs. We note that the individual demand is by construction a pure strategy as preferences are strict, see details below. 

We let $\chi_{s}^\theta=1$ denote that students of type $\theta$ demand school $s$ without deception, otherwise $\chi_{s}^\theta=0$ they do not. Such demand  requires first that the student must be eligible at $s$ and must prefer admission there ahead of the null match, which is captured in the first part of \eqref{eq:admission_eligibility}. In addition, the student must have no better options at other schools, which requires that $\hat{\chi}_{s,s'}^\theta=1$ and holds if either admission is not feasible or not profitable if feasible, see \eqref{eq:admission_eligibiliy_outside}.

We proceed with formally defining optimal deception and auxiliary measures, giving structure to the less rigorous definition from Section~\ref{sec:theory}. We let $\delta_{s}^\theta=1$ denote that student type $\theta$ demands school $s$ with deception and otherwise not $\delta_{s}^\theta=0$.
Demand with deception requires that a number of conditions hold, similar to demand without deception. First, the student must be ineligible for admission to $s$, but able to gain admission through deception, and must prefer admission at $s$ net of application costs ahead of the null match, which is captured in the first part of \eqref{eq:admission_deception}. Moreover, the student must have no better options at other schools, which requires that $\hat{\delta}_{s,s'}^\theta=1$ and holds if either admission is not feasible or not profitable if feasible, see \eqref{eq:admission_deception_outside}. We now denote the optimal deception for type $\delta^\theta=\max_{s\in\mathcal{S}}\delta_{s}^\theta$. We denote the aggregate optimal deception by $\delta=\int \delta^\theta(P,w) \eta(\theta) d\theta$. 

\begin{align}
\chi_{\theta,s}(P)&=\textbf{1}\big(v_{\theta,s}>0\big) \cdot \textbf{1}\big(p_{\theta,s}(a_{\theta}^{0})\ge P_s\big)\cdot \prod_{\forall s'\in(\mathcal{C}\backslash c)}\hat{\chi}_{s,s'}^\theta\label{eq:admission_eligibility}\\  
\delta_{\theta,s}(P)&=\textbf{1}\big(v_{\theta,s}>\gamma\big) \cdot \textbf{1}\big(p_{\theta,s}(a_{\theta}^{0})<P_s\big)\cdot \textbf{1}\big(p_{\theta,s}(a_{\theta,s}^{*})\ge P_s\big)\cdot \prod_{\forall s'\in(\mathcal{C}\backslash c)}\hat{\delta}_{s,s'}^\theta\label{eq:admission_deception}\\
\hat{\chi}_{s,s'}^\theta(P)&=
\begin{cases}
1 & \text{if either i) }v_{\theta,s}>v_{\theta,s'}\text{ or ii) } \max(p_{\theta,s}(a_{\theta}^{0}),p_{\theta,s}(a_{\theta,s}^{*}))<P_{s'}\\
 & \text{or iii) }\big(p_{\theta,s}(a_{\theta}^{0})<P_{s'}\text{ and }v_{\theta,s}>v_{\theta,s'}-\gamma\big)\\
0 & \text{otherwise}
\end{cases}\label{eq:admission_eligibiliy_outside}\\
\hat{\delta}_{s,s'}^\theta(P)&=
\begin{cases}
1 & \text{if either i) }v_{\theta,s}-\gamma>v_{\theta,s'}\text{ or ii) } \max(p_{\theta,s}(a_{\theta}^{0}),p_{\theta,s}(a_{\theta,s}^{*}))<P_{s'}\\
 & \text{or iii) }\big(p_{\theta,s}(a_{\theta}^{0})<P_{s'}\text{ and }v_{\theta,s}>v_{\theta,s'}\big)\\
0 & \text{otherwise}
\end{cases}\label{eq:admission_deception_outside}
\end{align}

We define the  \textbf{aggregate demand} for each school $s$ over students as:

\begin{align}
    D_s(P,w)=\int\eta(\theta)\,\textbf{1}[D^\theta(P,w)=s]\,\text{d}\theta
\end{align}

Let $D$ denote the vector of aggregate demand over the schools in $\mathcal{S}$. In addition, let $D_{\theta,s}=1$ if $D^\theta=s$, otherwise $D_{\theta,s}=0$.

\subsubsection{Properties of demand and optimal deception}\label{app:demand_properties}
A central property that \citet{azevedo2016supply} use to derive results about matching is that individual demand adjusts to admission cutoffs in the same way that price affects demand in standard consumer choice. One remarkable fact about our matching framework, which extends \citet{azevedo2016supply} to two kinds of blocking pairs, is that the same fundamental properties of demand about individuals are preserved. The two properties are defined as follows. We say that individual demand is \textbf{monotone} if $D^\theta_s$ is non-increasing in $P_s$ for any type. Individual demand satisfies \textbf{gross substitutes} if $D^\theta_s$ is non-decreasing in $P_{s'}$ for any $s'\ne s$.

\begin{figure}
    \centering
    \begin{subfigure}{.47\linewidth}
        \begin{tikzpicture}
    \draw[thick,->] (0,0) -- (5,0);
    \draw[thick,->] (0,0) -- (0,5);
    \draw (5.35, 0) node {$P_1$};
    \draw (0, 5.35) node {$P_2$};
    \draw (-.3, 0) node {$0$};
    \draw (0, -.3) node {$0$};
    \draw (-.3, 4) node {$1$};
    \draw (4, -.3) node {$1$};
    \draw (-.5, 1) node {$p^{exog}_{\theta,2}$};
    \draw (1, -.35) node {$p^{exog}_{\theta,1}$};
    \draw[fill=red!50] (0,4) -- (0,1)  -- (1,1) -- (1,4) -- cycle;
    \draw[fill=blue!50] (0,0) --  (0,1) -- (4,1) -- (4,0) -- cycle;
    \draw[fill=white] (1,1) -- (4,1) -- (4,4) -- (1,4) -- cycle;
    \draw (2.75, 2.75) node {$(\emptyset,0)$};
    \draw (.5, 2.75) node {$(1,0)$};
    \draw (2.25, .5) node {$(2,0)$};
\end{tikzpicture}
        \caption{No manipulability  ($w=0$)}
    \end{subfigure}%
    \begin{subfigure}{.49\linewidth}
        \begin{tikzpicture}
    \draw[thick,->] (0,0) -- (5,0);
    \draw[thick,->] (0,0) -- (0,5);
    \draw (5.35, 0) node {$P_1$};
    \draw (0, 5.35) node {$P_2$};
    \draw (-.3, 0) node {$0$};
    \draw (0, -.3) node {$0$};
    \draw (-.3, 4) node {$1$};
    \draw (4, -.3) node {$1$};
    \draw (-.5, 1) node {$p^{exog}_{\theta,2}$};
    \draw (1, -.35) node {$p^{exog}_{\theta,1}$};
    \draw (-1.05, 2) node {$p^{manip}_{\theta,2}(a_{\theta,2}^{*})$};
    \draw[fill=blue!20] (0,1) --  (0,2) -- (4,2) -- (4,1)-- cycle;
    \draw[fill=red!50] (0,4) -- (0,2)  -- (1,2) -- (1,4) -- cycle;
    \draw[fill=blue!50] (0,0) -- (0,1) -- (4,1) -- (4,0) -- cycle;
    \draw[fill=white] (1,2) -- (4,2) -- (4,4) -- (1,4) -- cycle;
    \draw (2.5, 3) node {$(\emptyset,0)$};
    \draw (.5, 3) node {$(1,0)$};
    \draw (2.5, .5) node {$(2,0)$};
    \draw (2.5, 1.5) node {$(2,1)$};
\end{tikzpicture}
        \caption{Complete manipulability ($w=1$)}
    \end{subfigure}%
    \caption{Individual demand and optimal deception
    \begin{minipage}{1\linewidth}
    \vspace{3mm}
    \footnotesize
    Notes: The panels depict individual demand and preparatory action under respectively no and complete manipulability of priority scores. The demand is depicted as a function of school cutoffs. The figure is based on a situation with a student of type $\theta$ and two schools denoted by 1,2. In the figure we assume that school 2 is preferred even  after subtracting manipulation costs: $v_{\theta,2}-\gamma>v_{\theta,1}$. The  elements of the tuple within the shaded areas denote respectively which school is demanded (most preferred among the feasible) and the optimal level of deception. We also assume that for both schools it holds that $p^{manip}_{\theta,s}(a_\theta^{0})=p^{exog}_{\theta,s}$. To illustrate this, consider the red shaded area in panel (a) with the tuple (1,0). In this area, the student is accepted at school 1 and she does not use deception. In the light-shaded blue area in panel (b) with the tuple (2,1), the student is accepted at school 2 and uses deception.
    \end{minipage}}
    \label{fig:demand_deception}
\end{figure}

To illustrate that demand satisfies these two properties we depict an example of demand  with and without deception in Figure~\ref{fig:demand_deception}. The figure shows that irrespective of deception or not, demand for each school is non-increasing in own cutoff and non-decreasing in the other school's cutoff. The result below generalizes how these propertíes hold more broadly.

\begin{lemma}\label{lemma:demand_grosssub}
At any cutoff it holds that individual demand is monotone and satisfies gross substitutes.
\end{lemma}

\noindent\textbf{Proof of Lemma~\ref{lemma:demand_grosssub}:} We focus on a given school $s$ and a given type $\theta$ and let $P_{-s}$ denote cutoffs of other schools. We will first prove that demand is \textit{monotone} in $P_s$, i.e. that $D^\theta_s(\cdot)$ is non-increasing in $P_s$. We see from \eqref{eq:admission_eligibility} and \eqref{eq:admission_deception} that demanding $s$ requires either $p_{\theta,s}(a_{\theta}^{0})\ge P_s$ or ($p_{\theta,s}(a_{\theta}^{0})<P_s$ and $p_{\theta,s}(a_{\theta,s}^{*})\ge P_s$). This means that type $\theta$ can only increase demand for school $s$ for a given $P_{-s}$ if $P_s$ is below a given threshold (=$\max\{p_{\theta,s}(a_{\theta}^{0}),p_{\theta,s}(a_{\theta,s}^{*})\}$). 

We proceed to demonstrate that demand satisfies \textit{gross substitutes}, i.e. $D^\theta_s(\cdot)$ is non-decreasing $P_{s'}$ where $s'\in(\mathcal{S}\backslash s)$. 
We see that type $\theta$ can only stop demanding $s$ (by starting to demand $s'$) if  \eqref{eq:admission_eligibiliy_outside} and \eqref{eq:admission_deception_outside} is violated, which requires that $p_{\theta,s'}(a_{\theta}^{0})>P_{s'}$ or ($p_{\theta,s'}(a_{\theta}^{0})<P_{s'}$ and $p_{\theta,s}(a_{\theta,s'}^{*})>P_{s'}$). In other words, type $\theta$ can only decrease demand for school $s$ keeping $P_{-s'}$ fixed if $P_{s'}$ is below some threshold (=$\min\{p_{\theta,s'}(a_{\theta}^{0}),p_{\theta,s'}(a_{\theta,s'}^{*})\}$). $\blacksquare$
\bigskip

An important consequence of the lemma is that all the properties of aggregate demand and matchings carry over from \citet{azevedo2016supply} to our setting.

We round off our analysis by proving Proposition~\ref{claim:monotone_deception} from the main text.
\bigskip

\noindent\textbf{Proof of Proposition~\ref{claim:monotone_deception}:} Let $P$ be fixed. We begin with part i). Suppose that the optimal level of deception decreases from $\delta^\theta_{w_0}=1$  to $\delta^\theta_{w_1}=0$ after an increase in the level of manipulability from $w_0$ to $w_1$. Let $s=D^\theta(P,w_1)$ and $s'=D^\theta(P,w_0)$. As $\delta^\theta_{w_0}=1$ it must be that $p_{\theta,s'}(a_{\theta,s'}^{*})\ge P_{s'}$ and $p_{\theta,s'}(a_{\theta}^{0})<P_{s'}$ under $w_0$ as otherwise admission at $s'$ would be possible without costly deception. This has the following implications: 
(i) under $w_0$, the priority score of $\theta$ at school $s'$ was increased by deception, i.e. $p_{\theta,s}(a_{\theta,s'}^{*})>p_{\theta,s'}(a_{\theta}^{0})$;
(ii) as $P$ is fixed it means that school $s$ was a feasible option under $w_0$ but less preferred than school $s'$ and thus that $u^\theta_{s'}-\gamma>u^\theta_{s}$. 
Moreover, as $w_1>w_0$ it follows that:

\begin{equation}
w_1\cdot p^{manip}_{i,s}(a_i)+(1-w_1)\cdot p^{exog}_{i,s}>w_0\cdot p^{manip}_{i,s}(a_i)+(1-w_0)\cdot p^{exog}_{i,s},  
\end{equation}
which means that school $s'$ is also feasible under $w_1$. However, this is not possible as $u^\theta_{s'}-\gamma>u^\theta_{s}$, which would mean that the agent was not rational.
This completes the proof of part i).

We proceed with part ii). Let $P$ be fixed. We begin with part i). Let $P$ be fixed. Suppose that the optimal level of deception increases from $\delta^\theta_{\gamma_0}=0$  to $\delta^\theta_{\gamma_1}=1$ after an increase in the cost of manipulability from $\gamma_0$ to $\gamma_1$. 
Let $s_0=D_0^\theta(P,w)$ under $\gamma_0$ and $s_1=D_1^\theta(P,w_0)$ under $\gamma_1$.
As $\delta^\theta_{\gamma_1}=1$ it must be that $p_{\theta,s'}(a_{\theta,s'}^{*})\ge P_{s'}$ and $p_{\theta,s'}(a_{\theta}^{0})<P_{s'}$  as otherwise admission at $s'$ would be possible without costly deception. However, this means that the deception under $\gamma_0$ will also to lead admission at $s'$. As $s$ was affordable without deception under $\gamma_1$ this implies that $u^\theta_{s'}-\gamma_1>u^\theta_{s}$. However, since $\gamma_1>\gamma_0$ this means that $u^\theta_{s'}-\gamma_0>u^\theta_{s'}-\gamma_1$ and thus $u^\theta_{s'}-\gamma_0>u^\theta_{s}$. This contradicts the assumption of rationality, which completes part ii).
$\blacksquare$
\bigskip

\subsection{Preference revelation equilibria in large economies}\label{app:large_economy}

We now outline a preference revelation game that captures school choice and possible deception. The students are the set of players and their utility is as outlined previously; they gain utility matching but deception is costly. Schools are considered passive actors following the procedure and specified priorities. A student's (combined) action consist of its preparatory action, $a^{\theta}$, and its rank order list, $L^{\theta}$. 
 We assume students know the exact distribution of other students' types, which makes it irrelevant to distinguish whether information is complete or incomplete.

For tractability, we focus only on large economies because under mild conditions on the distribution of agent types it holds that there is a unique equilibrium.  We consider a continuum $E$ economy where the number of students have unit mass. We maintain that the set of schools $\mathcal{S}$ is finite and thus, each school $s$ has a mass of available capacity of seats available denoted by $C_s$.  We let the student type be a random variable that follows a joint probability function, $\eta$, over the type space $\Theta$. We assume that $\eta$ is atomless, and thus, the measure of students who are indifferent about any two schools is zero. 

The \textbf{equilibrium} of the preference revelation game requires that students best respond to other agents' actions. We note that an agent's demand captures its best response in terms of submitted preferences and deception behavior for any given cutoffs and level of manipulability, see Appendix~\ref{app:demand_formal}. Therefore, we can limit our analysis to demand functions and cutoffs.

A market clearing cutoff is a vector of cutoffs that clears supply of and demand for schools.

\begin{definition}
Cutoffs $P$ are market clearing if they satisfy the following market clearing conditions:
\begin{itemize}
    \item $D_s(P)\le C_s$ for all schools
    \item $D_s(P)= C_s$ for every school where $P_s>-\infty$
\end{itemize}
\end{definition}

We are ready to demonstrate our result on equilibrium uniqueness. In the remainder of this subsection we denote the dependence of the demand on measures as $D(\cdot|\eta)$. To express our result we need a final definition of smoothness for our measure:

\begin{definition}The distribution of student types $\eta$ is regular if the image under $D(\cdot|\eta)$ of the closure of the set \[\{P\in\mathbb{R}^\mathcal{S}: \,D(\cdot|\eta) \mbox{ is continuously differentiable at }P\}\] has Lebesgue measure 0.\end{definition}

Examples of sufficient conditions for regularity of $\eta$ includes $D(\cdot|\eta)$ being continuously differentiable or if $\eta$ has a continuous density. The result below states sufficient conditions for there to be a unique stable matching in our setting.\footnote{We note that Theorem~\ref{theorem:unique_equilibrium} is identical to \citet{azevedo2016supply} and that our proof demonstrates how their results can be generalized and extended to our setting.}

\begin{theorem}\label{theorem:unique_equilibrium}
Consider an economy $E = [\eta, C]$.\begin{enumerate}
    \item If $\eta$ has full support, then preference revelation game of the economy $E$ has a unique equilibrium.
    \item If $\eta$ is any regular distribution, then for almost every vector of capacities $C$ with $\sum_{S\in \mathcal{S}}C_s < 1$, then the preference revelation game of the economy $E$ has a unique equilibrium.
\end{enumerate}\end{theorem}

\noindent\textbf{Proof for Theorem~\ref{theorem:unique_equilibrium}:} 
To establish our proof we argue that the essential conditions for establishing the results in \citet{azevedo2016supply} are fulfilled. This proof is adapted from \cite{Bjerre-Nielsen2021VoluntaryInformation}

We begin by leveraging  Lemma~\ref{lemma:demand_grosssub}, which establishes that the individual demand function does indeed satisfy monotonicity and gross substitutes. Therefore, we may repeat the steps of \citep{azevedo2016supply} in deriving the same properties they only require properties of demand. In the following these properties in references to results in \citep{azevedo2016supply} are listed: First, to show aggregate demand is monotone and satisfy gross-substitutes (Remark A1). Second, construct an auxiliary mapping from cutoffs to market clearing cutoffs and use this to show existence of stable matchings and that the set of cutoffs is a lattice (respectively Corollary A1 and Theorem A1). Third, use the lattice structure to establish that there are is a smallest and largest cutoff (Proposition A2) as well as showing establishing a rural hospital theorem (Theorem A2), implying that the measure of students matched to each colleges is the same across equilibria.

To establish to establish the first part of Theorem~\ref{theorem:unique_equilibrium} on uniqueness given that $\eta$ has full support we note that the proof for Theorem 1, part 1 in \citep{azevedo2016supply} does not require anything beyond full support, which can also be assumed in our setting. 

To establish to establish the second part, we only note that requirements of regularity put restriction on the aggregate demand for given schools conditional on the measure $\nu$. However, these properties of aggregate demand either follow directly from results already established or from the assumption of regularity.  $\blacksquare$
\bigskip

We round off the theoretical analysis by how the level manipulability, $w$, affects aggregate deception behavior. We assume that $\eta$ is smooth in $\theta$, which allows us to examine marginal changes of $w$.

\begin{proposition}
\label{claim:marginal_deception_agg}
The total marginal effect on aggregate deception from changes in the level of manipulability ($w$) or the cost of manipulability ($\gamma$) can be decomposed into:
\begin{align}\label{eq:aggregate_deception_manipulability}
\frac{d}{d w}\hat{X}&=\frac{\partial}{\partial w}\hat{X}+\sum_{s\in \mathcal{S}}\left(\frac{\partial P_s}{\partial w}\frac{\partial}{\partial P_s}\hat{X}\right)\\
\frac{d}{d \gamma}\hat{X}&=\frac{\partial}{\partial \gamma}\hat{X}+\sum_{s\in \mathcal{S}}\left(\frac{\partial P_s}{\partial \gamma}\frac{\partial}{\partial P_s}\hat{X}\right)
\end{align}
where the direct effect of the manipulability level is strictly positive (i.e., $\frac{\partial}{\partial w}\hat{X}>0$) and the direct effect of the cost of manipulability is strictly negative (i.e., $\frac{\partial}{\partial \gamma}\hat{X}>0$).
\end{proposition}

\noindent\textbf{Proof of Proposition~\ref{claim:marginal_deception_agg}:} 
We begin proving the property of the total marginal effect from changes to the level of manipulability. We define the aggregate deception as a function of $w$ in the following way:

\begin{equation}
\hat{X}=\int x^\theta(P^*(w),w) \eta(\theta)d\theta
\end{equation}

Following Leibniz' rule, the aggregate deception is differentiable in $w$ and can be computed as the term below as the integral is indefinite:

\begin{align*}
\frac{d}{dw}\hat{X}
&=\int \frac{d}{dw}x^\theta(P^*(w),w) \eta(\theta)d\theta\\
&=\int \frac{\partial}{\partial w}x^\theta(P^*(w),w) \eta(\theta) d\theta+\sum_{s\in\mathcal{S}} \int \frac{\partial P_s}{\partial w}\frac{\partial}{\partial P_s}x^\theta(P^*(w),w) \eta(\theta) d\theta\\
&=\frac{\partial}{\partial w}\hat{X}+\sum_{s\in \mathcal{S}}\left(\frac{\partial P_s}{\partial w}\frac{\partial}{\partial P_s}\hat{X}\right)
\end{align*}

We now finalize the proof by showing that the direct effect, $\frac{\partial}{\partial w}\hat{X}$, is strictly positive. This follows from the assumption of full support and by applying Proposition~\ref{claim:monotone_deception}.

The approach to proving the property of the total marginal effect from changes to the cost of manipulability is identical to the ones above. The only difference is that sign of the direct effect is strictly negative as this property is established in Proposition~\ref{claim:monotone_deception}. 
$\blacksquare$
\bigskip

The above proposition decomposes the effects into a direct effect and a general equilibrium effect. The direct effect is always positive for changes to the level of manipulability (reverse for the cost of manipulability), which means that when disregarding the general equilibrium the level of manipulability increases deception.

   \clearpage
\FloatBarrier
    \renewcommand{\thetable}{B.\arabic{table}}
    \renewcommand{\thefigure}{B.\arabic{figure}}
    \renewcommand{\setthesection}{\Alph{section}}
    \renewcommand{\setthesubsection}{\Alph{subsection}}
    \renewcommand{\thetable}{B.\arabic{table}}
    \renewcommand{\thefigure}{B.\arabic{figure}}
\section{Policy changes appendix}
\begin{figure}[h!]
\includegraphics[width=1\linewidth]{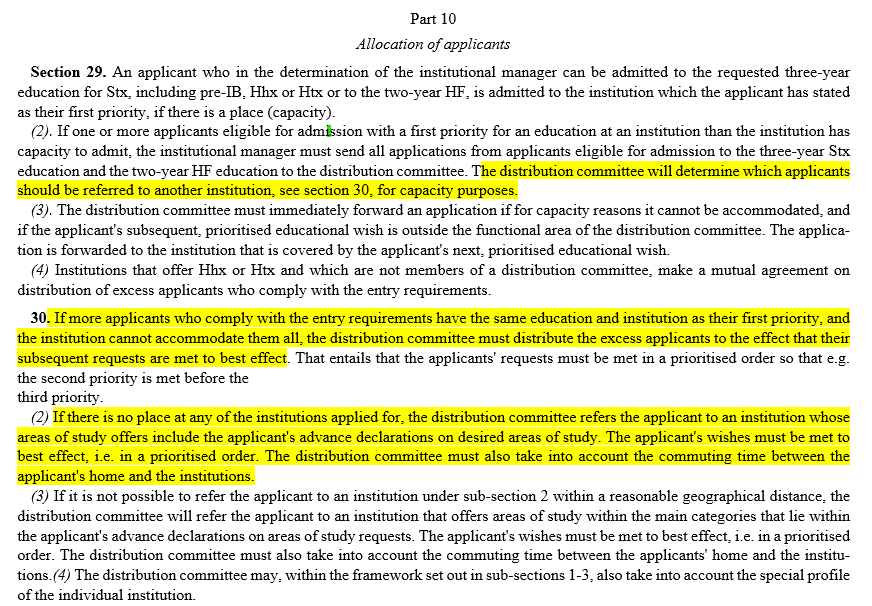}
 \caption{Pre 2012 assignment law. Translated from:  \href{https://www.retsinformation.dk/eli/lta/2009/694}{https://www.retsinformation.dk/eli/lta/2009/694}. }
\label{fig:pre2012law}
\end{figure}

\begin{figure}[h!]
\includegraphics[width=1\linewidth]{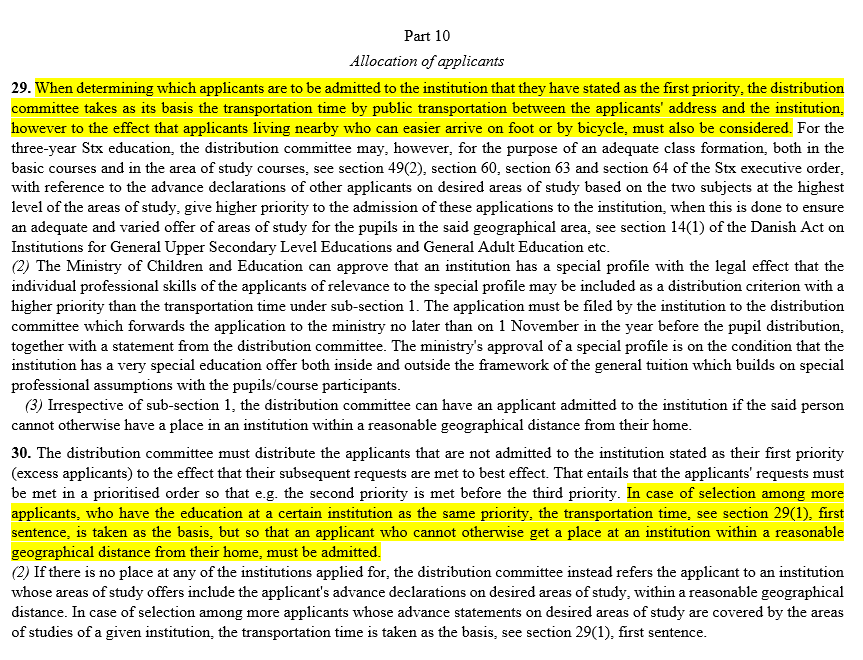}\\
 \caption{Post 2011 assignment law. Translated from:  \href{https://www.retsinformation.dk/eli/lta/2012/174}{https://www.retsinformation.dk/eli/lta/2012/174}.}
\label{fig:post2012law}
\end{figure}

\clearpage
\FloatBarrier
\section{Additional figures and tables}
    \renewcommand{\thetable}{C.\arabic{table}}
    \renewcommand{\thefigure}{C.\arabic{figure}}
    \renewcommand{\setthesection}{\Alph{section}}
    \renewcommand{\setthesubsection}{\Alph{subsection}}
    \renewcommand{\thetable}{C.\arabic{table}}
    \renewcommand{\thefigure}{C.\arabic{figure}}

\begin{figure}[h!]
\subfloat[Individual address changes\label{individual}]{\includegraphics[width=.9\linewidth]{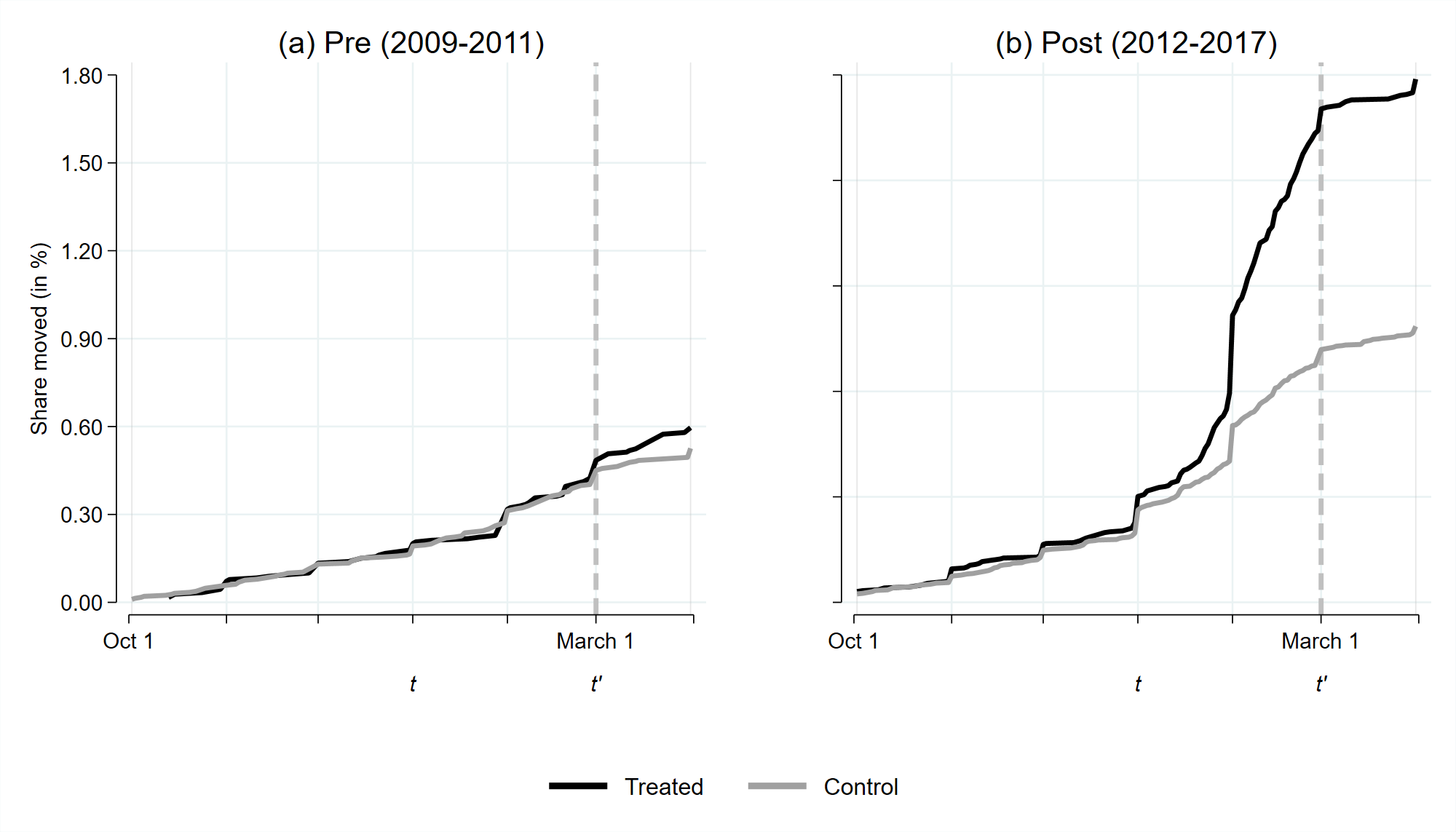}}\\
\subfloat[Household address changes\label{household}]{\includegraphics[width=.9\linewidth]{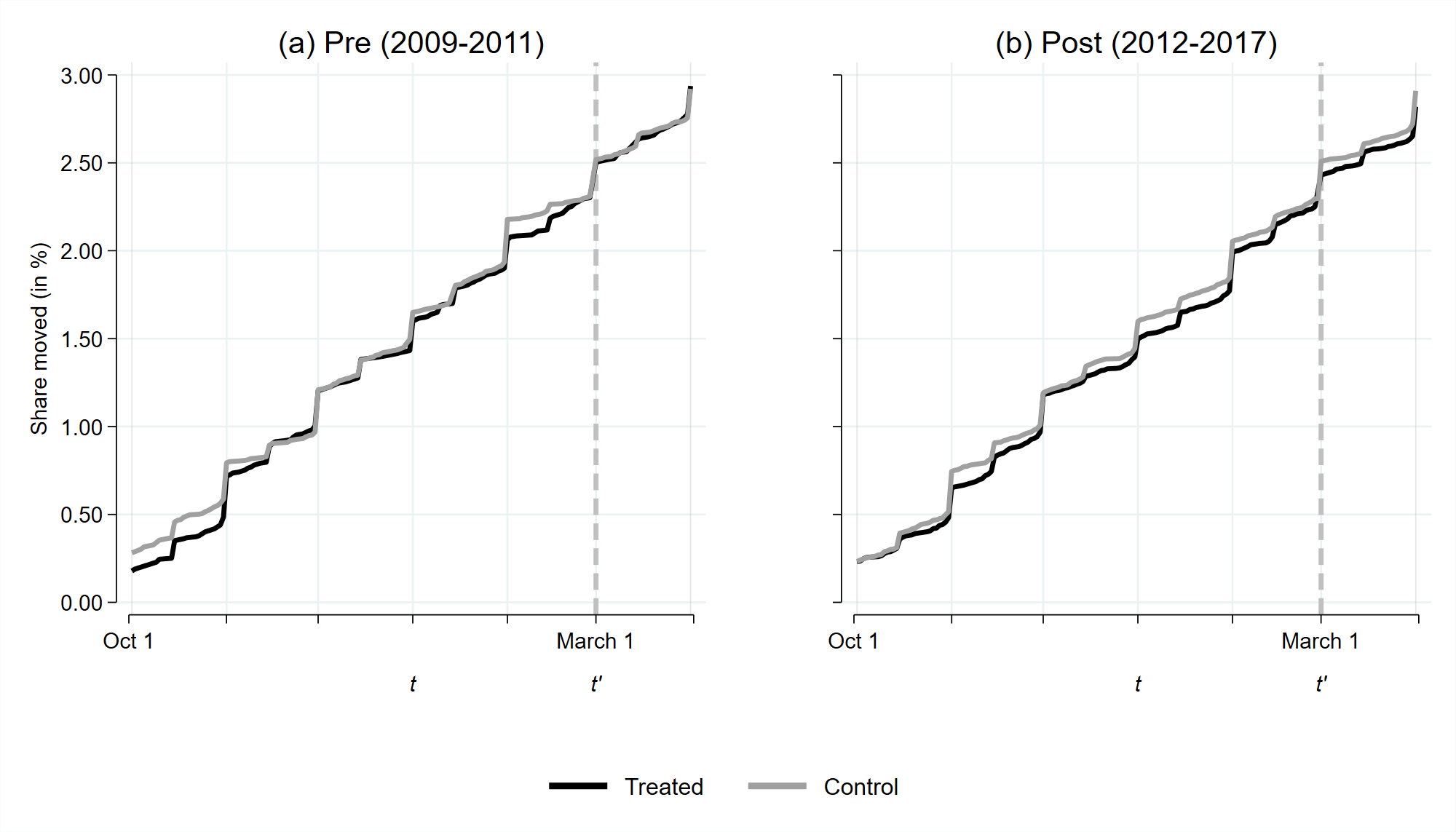}}
 \caption{Cumulative individual and household address changes, October 1 to April 1 \vspace{6pt}  \\ 
\begin{minipage}{1\linewidth}
\footnotesize
Notes: The Figure shows cumulative   daily address changes relative to the initial population for all students in the last year of compulsory schooling, who applied for a high school. In Figure \ref{individual} we only show address changes where no other member of the household changed the address. In Figure \ref{household} we only show address changes where at least one  other member of the household changed the address.
\end{minipage}}
\label{fig:ab}
\end{figure}

\begin{figure}[h!]
\subfloat[By application year]{\includegraphics[width=.5\linewidth]{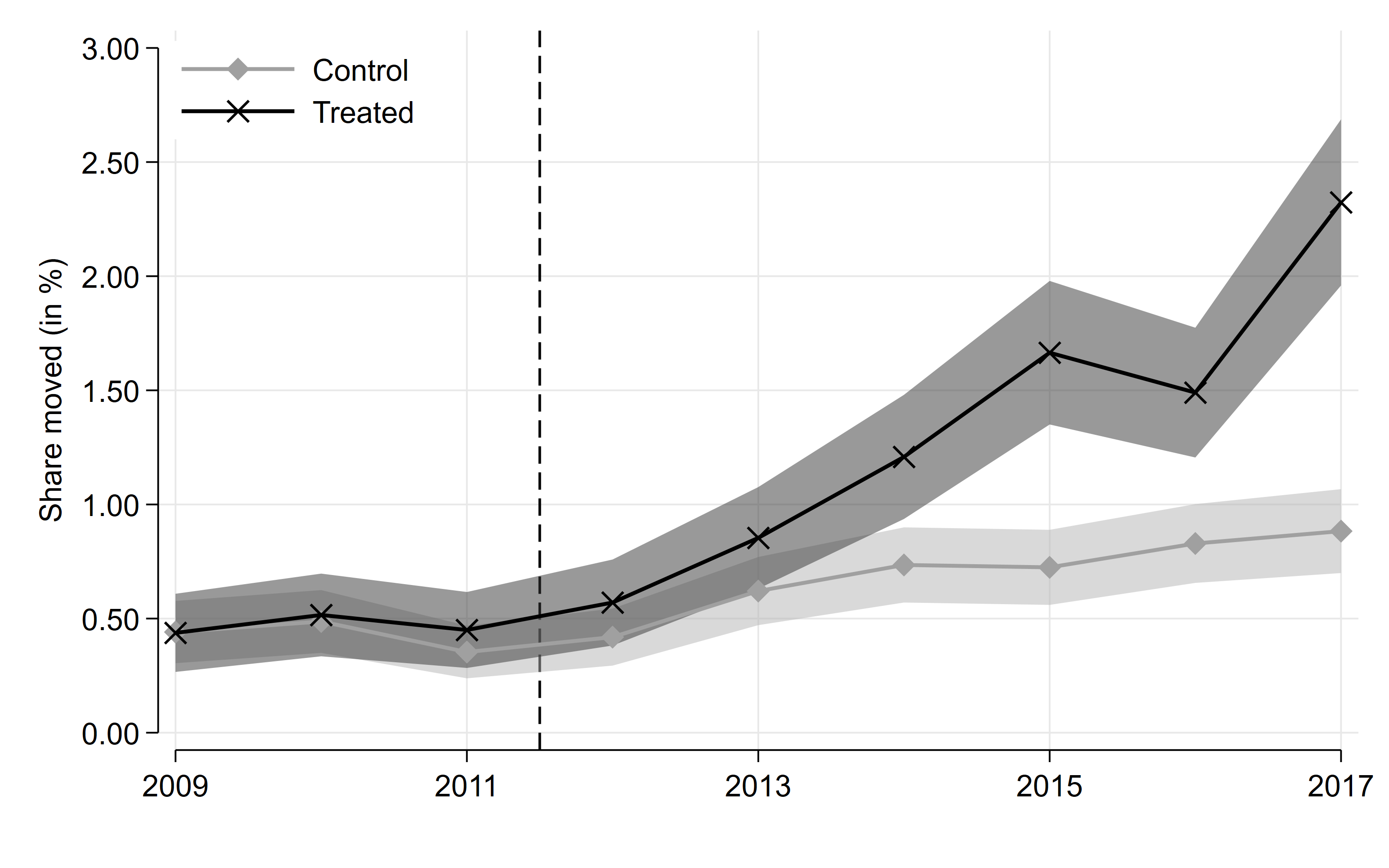}}
\subfloat[By enrollment]{\includegraphics[width=.5\linewidth]{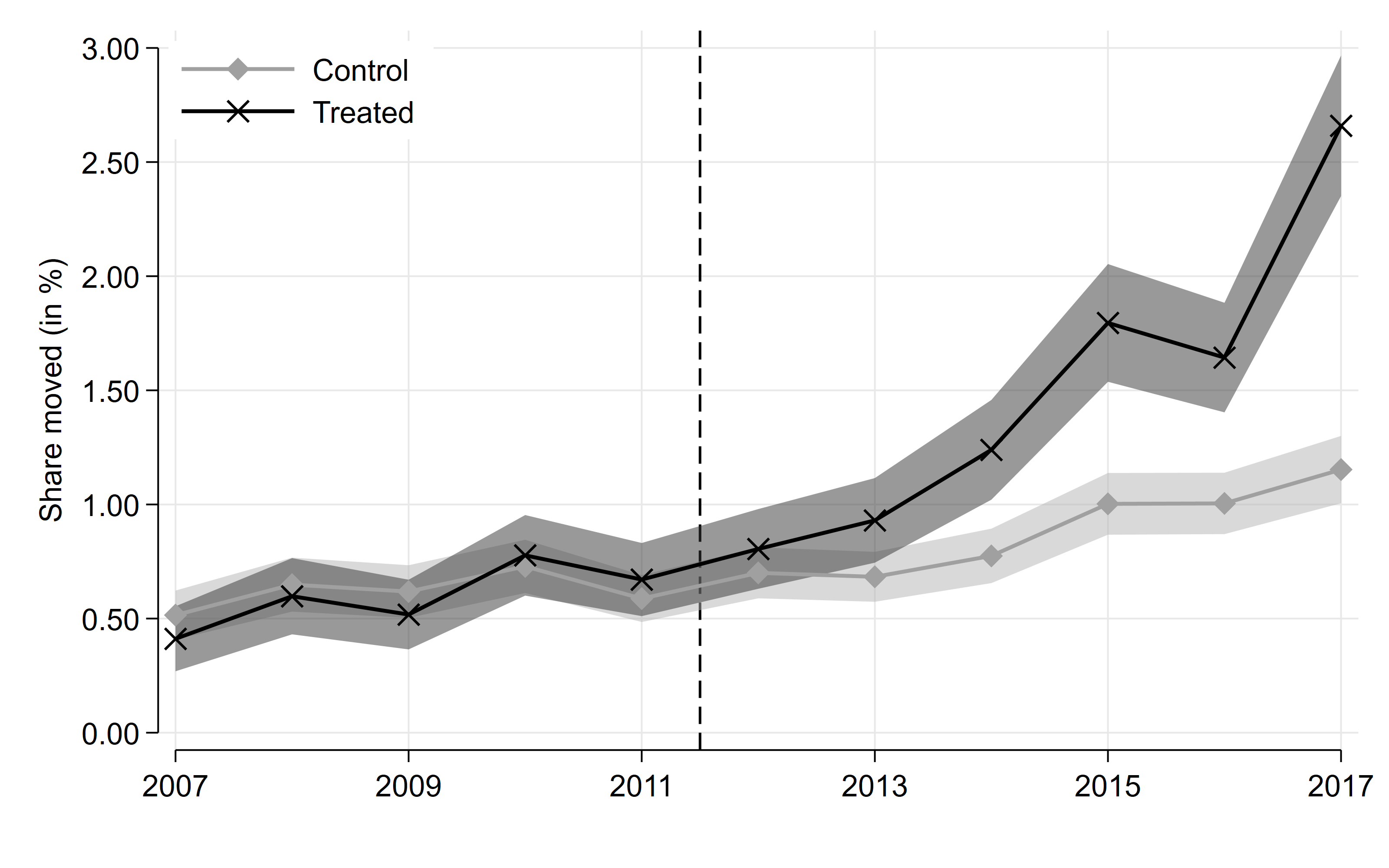}}
\caption{Address changes between October 1 and March 1 over time\vspace{6pt} \\ 
\begin{minipage}{1\linewidth}
\footnotesize
Notes: The shows the share of students who changed their address between October 1 and March 1 in respectively, the year of high school application and in the year of high school enrollment. 
\end{minipage}}
\label{fig:trends}
\end{figure}

\begin{figure}[h!]
\includegraphics[width=\linewidth]{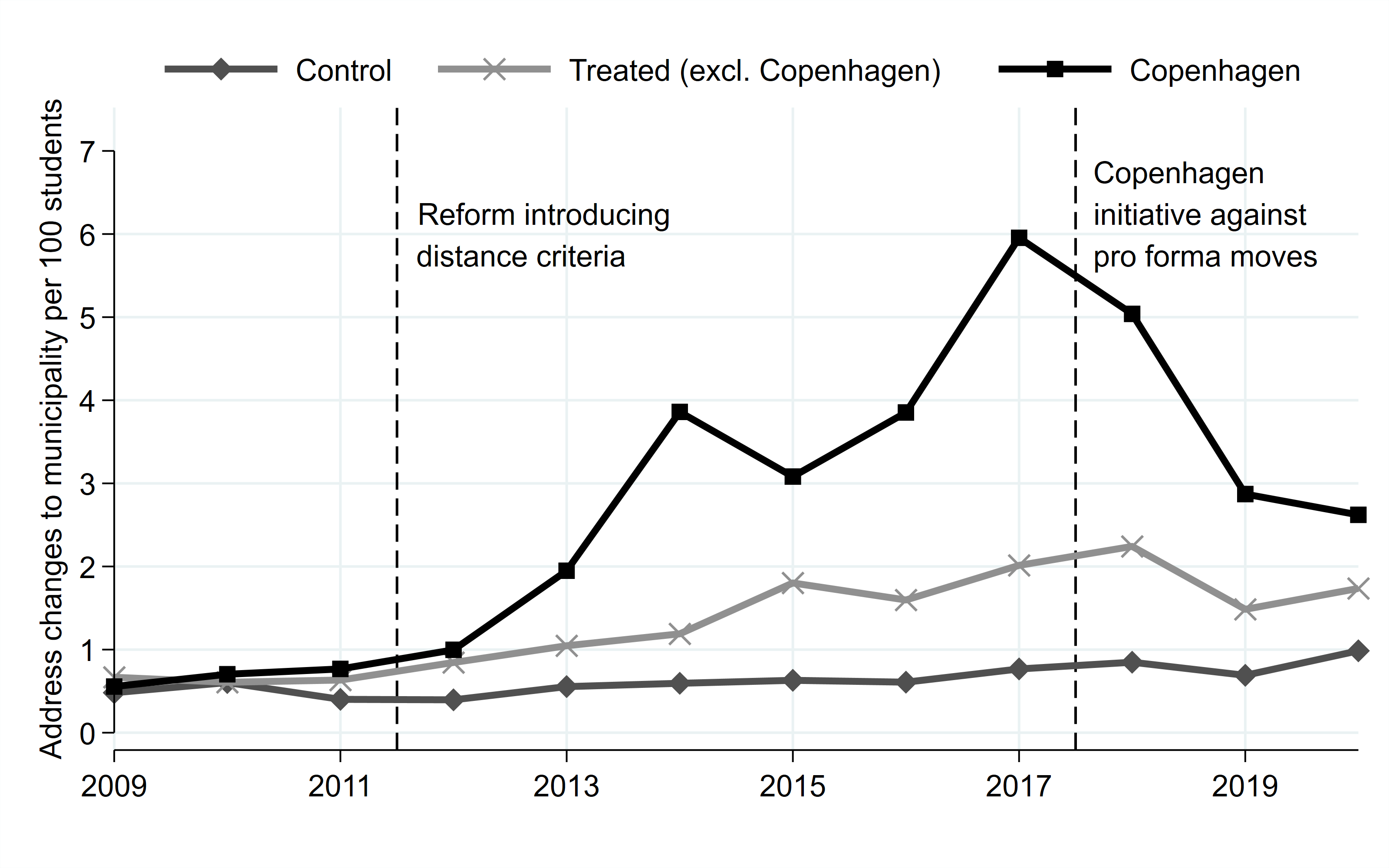}
\caption{Address changes between October 1 and March 1 over time\vspace{6pt} \\ 
\begin{minipage}{1\linewidth}
\footnotesize
Notes: The shows the share of students who changed their address between October 1 and March 1 in the last year before high school enrollment for the treated and control groups. Only address changes where no other member of the household change the address are considered. 
\end{minipage}}
\label{fig:trendsl4_appendix}
\end{figure}

\begin{figure}[h!]
\includegraphics[width=.95\linewidth]{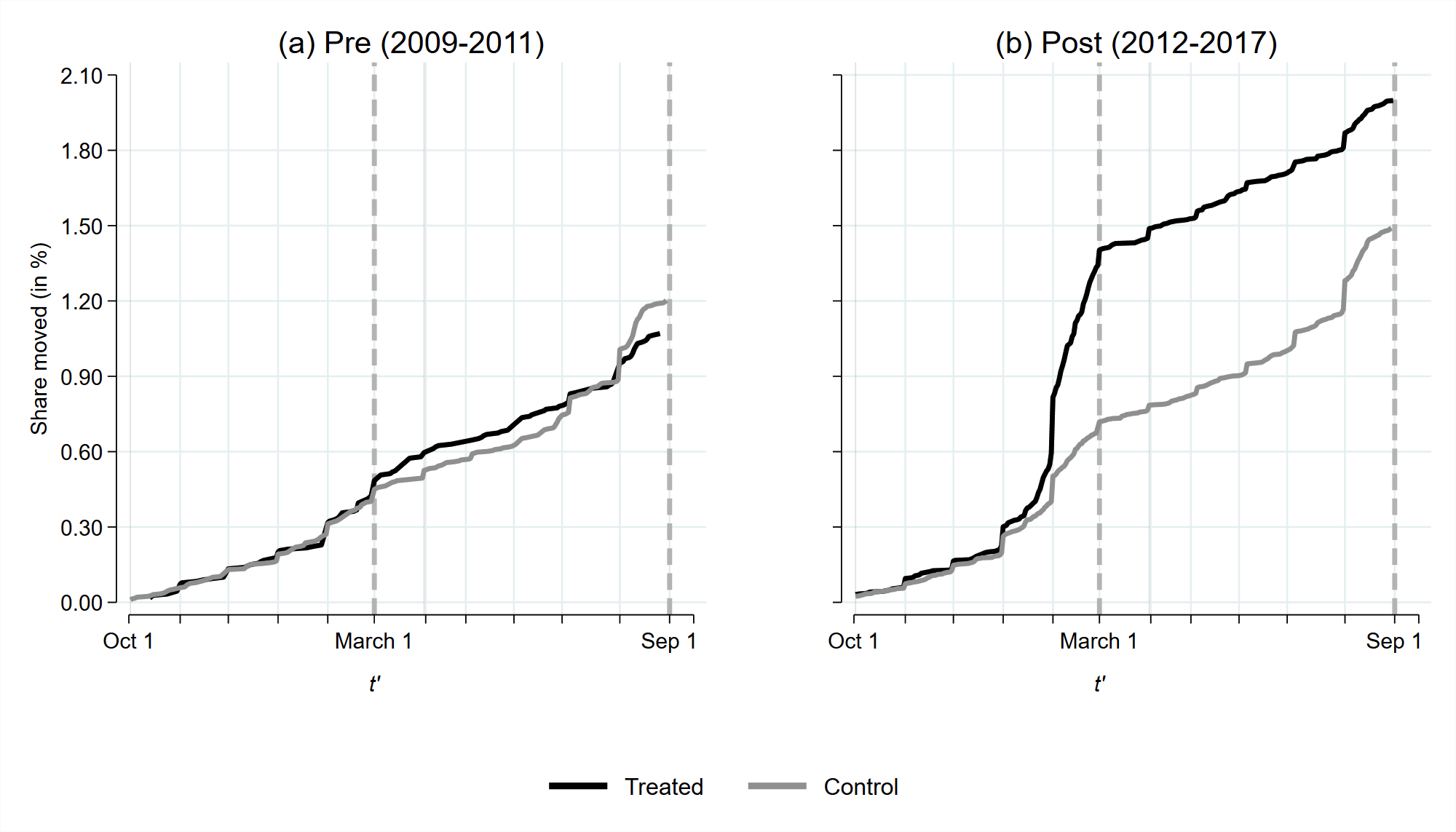}
\caption{Cumulative address changes, October 1 to August 1 \vspace{6pt} \\ 
\begin{minipage}{1\linewidth}
\footnotesize
Notes: The Figure shows cumulative   daily address changes relative to the initial population for all students in the last year of compulsory schooling, who applied for a high school. 
\end{minipage}}
\label{fig:addressmanipulationfull}
\end{figure}

\begin{table}[h!]
    \caption{Regression results: The effect of high-school application incentives on moving behavior, individual address changes only - considering all individuals finishing compulsory schooling.}
    \label{tab:res9}
    \begin{tabularx}{1\linewidth}{cXc cc cc ccc}
    \toprule
   &&\multirow{4}{*}{All}&&&\multicolumn{4}{c}{Parental income}\\
    &&&&&\multicolumn{2}{c}{$\leq$p50}&\multicolumn{2}{c}{$>$p50}\\
    &&&\multicolumn{2}{c}{Parental income}&\multicolumn{4}{c}{Middle school GPA}\\
    &&&$\leq$p50&$>$p50&$\leq$p50&$>$p50&$\leq$p50&$>$p50\\
    &&(1)&(2)&(3)&(4)&(5)&(6)&(7)\\
    \midrule
   &Post $\times$ treated&       0.003&       0.001&       0.006&      -0.001&       0.004&       0.002&       0.008\\
&                    &     (0.002)&     (0.002)&     (0.003)&     (0.002)&     (0.003)&     (0.003)&     (0.003)\\
\midrule
&Observations        &     703,926&     351,649&     350,870&     206,673&     118,353&     127,060&     212,907\\
&MDV                 &       0.041&       0.053&       0.028&       0.058&       0.037&       0.034&       0.023\\[-12pt]
\\
    \bottomrule
    \end{tabularx}\\
      \begin{minipage}{1\linewidth}
           \footnotesize Notes: The table shows the coefficients from estimating equation \eqref{eq:1} using ordinary least squares. The dependent variable takes the value of 1 if the individual changed the address before March 1, and 0 otherwise.  The variable treated is 1 for individuals living in municipalities where demand for high schools is in the top quartile (see Figure \ref{fig:geographvara} (a)) and zero otherwise. The variable post is one in all years after 2011. All regressions include the full set of year and municipality fixed effects, as well as the  full set of controls. The full set of controls include a variable for the average parental years of schooling, a variable for the average parental disposable income, a variable for the 9th grade GPA, and an indicator for the biological gender being female. We also include indicator variables for missing parental education, missing parental income, and missing 9th grade GPA. The columns (2) to (7) show sub-sample regressions based on parental disposable income and middle school GPA, for both measures the median is calculated by cohort and municipality.         MDV is the mean of the dependent variable for untreated individuals.         Standard errors clustered at the municipality of residence at the beginning of the last year before high school enrollment in parenthesis. 
    \end{minipage}
\end{table}

\begin{table}[h!]
    \caption{Regression results: Extensive vs intensive margin response, individual address changes only}
    \label{tab:extensive}
    \begin{tabularx}{0.95\linewidth}{Xc cc cc ccc}
    \toprule
   &\multirow{4}{*}{All}&&&\multicolumn{4}{c}{Parental income}\\
    &&&&\multicolumn{2}{c}{$\leq$p50}&\multicolumn{2}{c}{$>$p50}\\
    &&\multicolumn{2}{c}{Parental income}&\multicolumn{2}{c}{middle school GPA}&\multicolumn{2}{c}{middle school GPA}\\
    &&$\leq$p50&$>$p50&$\leq$p50&$>$p50&$\leq$p50&$>$p50\\
    &(1)&(2)&(3)&(4)&(5)&(6)&(7)\\
    \toprule
    Total                  &       0.006 &       0.003 &       0.009 &       0.003  &       0.004         &       0.008 &       0.009 \\
                    &     (0.002)         &     (0.001)         &     (0.004)         &     (0.001)         &     (0.003)         &     (0.003)         &     (0.004)         \\
Extensive                  &       0.006 &       0.002         &       0.010 &       0.000         &       0.003         &       0.006  &       0.012 \\
                    &     (0.002)         &     (0.002)         &     (0.003)         &     (0.003)         &     (0.003)         &     (0.003)         &     (0.004)         \\
Intensive                  &       0.000         &       0.001         &      -0.001         &       0.002         &       0.001         &       0.002         &      -0.003 \\
                    &     (0.001)         &     (0.001)         &     (0.001)         &     (0.002)         &     (0.002)         &     (0.002)         &     (0.001)         \\
MDV                 &       0.006         &       0.006         &       0.006         &       0.006         &       0.006         &       0.006         &       0.007         \\
\midrule
Observations        &     147,903         &      74,155         &      73,674         &      44,757         &      28,751         &      29,080         &      44,012         \\\\[-12pt]
    \midrule
    \end{tabularx}\\
    \begin{minipage}{0.95\linewidth}
           \footnotesize Notes: The table shows the coefficients from estimating equation \eqref{eq:1} using ordinary least squares. The dependent variable takes the value of 1 if the individual changed the address before March 1, and 0 otherwise.  The variable treated is 1 for individuals living in municipalities where demand for high schools is in the top quartile (see Figure \ref{fig:geographvara} (a)) and zero otherwise. The variable post is one in all years after 2011. All regressions include the full set of year and municipality fixed effects, as well as the  full set of controls. The full set of controls include a variable for the average parental years of schooling, a variable for the average parental disposable income, a variable for the 9th grade GPA, and an indicator for the biological gender being female. We also include indicator variables for missing parental education, missing parental income, and missing 9th grade GPA.  The total response is the estimated coefficient for $\beta$  considering address changes between October 1 and March 1. The Extensive response is the estimated coefficient on $\beta_3$ considering all address changes between October 1 and August 1. The Intensive response is the difference between these two coefficients.  The total and extensive response are found by estimating equation \eqref{eq:1} using ordinary least squares.  Only individual address changes are considered. MDV is the mean of the dependent variable for the untreated individuals. Bootstrapped standard errors using 200 iterations and clustered at the municipality of residence at the beginning of the last year before high school enrollment in parenthesis.
    \end{minipage}
\end{table}

\begin{table}[h!]
    \caption{Regression results: Address change to relative}
    \label{tab:relative}
    \begin{tabularx}{0.95\linewidth}{Xc cc cc ccc}
    \toprule
   &\multirow{4}{*}{All}&&&\multicolumn{4}{c}{Parental income}\\
    &&&&\multicolumn{2}{c}{$\leq$p50}&\multicolumn{2}{c}{$>$p50}\\
    &&\multicolumn{2}{c}{Parental income}&\multicolumn{2}{c}{middle school GPA}&\multicolumn{2}{c}{middle school GPA}\\
    &&$\leq$p50&$>$p50&$\leq$p50&$>$p50&$\leq$p50&$>$p50\\
    &(1)&(2)&(3)&(4)&(5)&(6)&(7)\\
    \toprule
    Total                  &       0.006 &       0.003 &       0.009 &       0.003  &       0.003         &       0.008 &       0.010 \\
                    &     (0.002)         &     (0.001)         &     (0.004)         &     (0.002)         &     (0.003)         &     (0.003)         &     (0.004)         \\
Not family                  &       0.003 &       0.002 &       0.004 &       0.003 &       0.001         &       0.002         &       0.005 \\
                    &     (0.001)         &     (0.001)         &     (0.002)         &     (0.001)         &     (0.002)         &     (0.002)         &     (0.002)         \\
Family                  &       0.003 &       0.001         &       0.005 &       0.000         &       0.002         &       0.006 &       0.005 \\
                    &     (0.001)         &     (0.001)         &     (0.002)         &     (0.001)         &     (0.002)         &     (0.002)         &     (0.002)         \\
\midrule
Observations        &      147,903         &       74,155         &       73,674         &       44,757         &       28,751         &       29,080         &       44,012         \\
\\[-12pt]
    \midrule
    \end{tabularx}\\
    \begin{minipage}{0.95\linewidth}
         \footnotesize Notes: The table shows the coefficients from estimating equation \eqref{eq:1} using ordinary least squares. The dependent variable takes the value of 1 if the individual changed the address before March 1, and 0 otherwise.  The variable treated is 1 for individuals living in municipalities where demand for high schools is in the top quartile (see Figure \ref{fig:geographvara} (a)) and zero otherwise. The variable post is one in all years after 2011. All regressions include the full set of year and municipality fixed effects, as well as the  full set of controls. The full set of controls include a variable for the average parental years of schooling, a variable for the average parental disposable income, a variable for the 9th grade GPA, and an indicator for the biological gender being female. We also include indicator variables for missing parental education, missing parental income, and missing 9th grade GPA. The total response is the estimated coefficient for $\beta$  considering address changes between October 1 and March 1. The Not Family coefficient is coefficient from the same regression, but excluding address changes to a family member.  The Family coefficient is the coefficient when only considering address changes to a family member. Family member is defined as the biological grandparents and the siblings of the biological parents.   Only individual address changes are considered. MDV is the mean of the dependent variable for the untreated individuals. Bootstrapped standard errors using 200 iterations and clustered at the municipality of residence at the beginning of the last year before high school enrollment in parenthesis. 
    \end{minipage}
\end{table}

\begin{figure}[h!]
\includegraphics[width=1\linewidth]{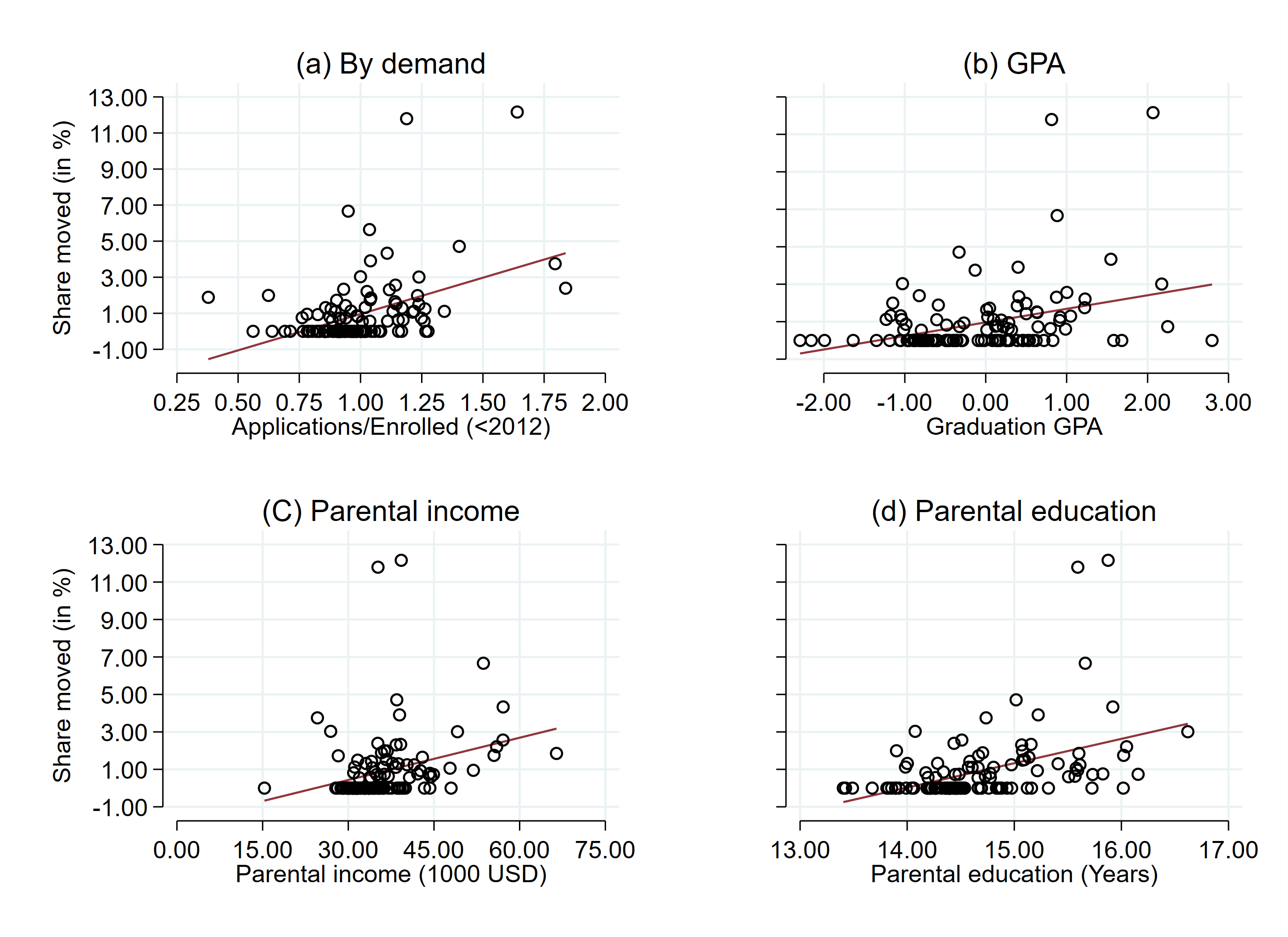}
\caption{High school level averages in address changes in 2017 \vspace{6pt} \\ 
\begin{minipage}{1\linewidth}
\footnotesize
Notes: Each circle shows the average for a high school. Graduation GPA, Parental Income, and Parental Education are measured for the graduates in the year before the share of moves is measured.  
\end{minipage}}
\label{fig:howcommonlevels}
\end{figure}

\begin{table}[ht!]
    \caption{Summary statistics, 2018 reform sample}
    \label{tab:sumstat2018}
    \begin{tabularx}{1\linewidth}{cXcc|ccccc}
        \toprule
    &&\multicolumn{2}{c}{--- All --- }&\multicolumn{3}{c}{Means for HS applicants }\\
    &&\multirow{2}{*}{Mean}&\multirow{2}{*}{SD}&\multirow{2}{*}{All}&\multicolumn{2}{c}{Over-subscribed }\\
    &&&&&No&Yes\\
            \midrule
        \multicolumn{6}{l}{A. Individual characteristics }\\
        &Applied to high school&  0.26&  0.44&  1.00&  1.00&  1.00\\
&Applied to oversubscribed high school&  0.47&  0.50&  0.47&  0.00&  1.00\\
&Enrolled in high school&  0.18&  0.39&  0.66&  0.91&  0.88\\
&Enrolled in oversubscribed high school&  0.42&  0.49&  0.43&  0.00&  0.92\\
&Enrolled in 1st priority high school&  0.94&  0.23&  0.83&  0.90&  0.80\\
\midrule
&Observations& 238,737& &  63,139&  24,696&  21,996\\
\\
        \multicolumn{6}{l}{B. High school application behaviour }\\
       \\[-12pt]
        
        \bottomrule
        \end{tabularx}
        \begin{minipage}{1\linewidth}
                        \footnotesize
        Notes: Parental schooling is the average among observed parents. Parental schooling is measured in years and is the average among observed parents. Both parental schooling and income are measured in the year of high school entry. A high school is defined as over-subscribed if more students listed the high school first in their application than the number of students that enrolled. 
        \end{minipage}
\end{table}

\begin{table}[h!]
    \caption{Regression results: By subgroups}
    \label{tab:simulationreg}
    \begin{tabularx}{0.95\linewidth}{cXc cc cc ccc}
    \toprule
    &&(1)&(2)&(3)&(4)\\
    \midrule
    &Post $\times$ treated&       0.000&       0.003&       0.007&       0.005\\
&                    &     (0.002)&     (0.002)&     (0.004)&     (0.003)\\
&Share moved         &       0.008&       0.011&       0.011&       0.016\\
&Share manipulative moves&       0.061&       0.235&       0.639&       0.306\\
&Observations        &      18,559&      32,506&       8,774&      13,160\\
\\[-12pt]
        &Female&No&Yes&No&Yes\\
    &Education$>$p50&No&No&Yes&Yes\\
    &Income$>$p50&No&No&No&No\\[12pt]
        \toprule
    &&(5)&(6)&(7)&(8)\\
    \midrule
   &Post $\times$ treated&       0.002&       0.008&       0.013&       0.010\\
&                    &     (0.004)&     (0.003)&     (0.004)&     (0.004)\\
&Share moved         &       0.011&       0.015&       0.019&       0.019\\
&Share manipulative moves&       0.214&       0.552&       0.696&       0.489\\
&Observations        &       9,602&      15,534&      21,363&      27,068\\\\[-12pt]
          &Female&No&Yes&No&Yes\\
     &Education$>$p50&No&No&Yes&Yes\\
     &Income$>$p50&Yes&Yes&Yes&Yes\\
    \midrule
    \end{tabularx}\\
    \begin{minipage}{0.95\linewidth}
        \footnotesize Notes: This table presents estimates for $\beta_3$ from equation \ref{eq:1} considering only individual address changes and estimated for eight subgroups by gender, parental income, and parental income. The dependent variable takes the value of 1 if the individual changed the address before March 1, and 0 otherwise. The variable treated is 1 for individuals living in municipalities where demand for high schools is in the top quartile (see Figure \ref{fig:geographvara} (a)) and zero otherwise. The variable post is one in all years after 2011. All regressions include the full set of year and municipality fixed effects, as well as the  full set of controls. The full set of controls include a variable for the average parental years of schooling, a variable for the average parental disposable income, a variable for the 9th grade GPA, and an indicator for the biological gender being female. We also include indicator variables for missing parental education, missing parental income, and missing 9th grade GPA. The share of manipulative moves is the ratio of the coefficient on the Post $\times$ treated variable relative to the share that moved.  Standard errors  clustered at the municipality of residence at the beginning of the last year before high school enrollment in parenthesis. 
    \end{minipage}
\end{table}

   \clearpage
\FloatBarrier
    \renewcommand{\thetable}{D.\arabic{table}}
    \renewcommand{\thefigure}{D.\arabic{figure}}
    \renewcommand{\setthesection}{\Alph{section}}
    \renewcommand{\setthesubsection}{\Alph{subsection}}
    \renewcommand{\thetable}{D.\arabic{table}}
    \renewcommand{\thefigure}{D.\arabic{figure}}

\section{Survey appendix}

\begin{table}[!h]\centering
	\def\sym#1{\ifmmode^{#1}\else\(^{#1}\)\fi}
	\caption{Wording of survey questions 
		\label{app:survey}}
		\begin{tabularx}{\textwidth}{ l X X }
			\hline\hline
			 Number&\multicolumn{1}{c}{Question} & \multicolumn{1}{c}{Scale} \\
			\hline
			Q1 & Have you changed your address formally during the last three months? With formal address change means a change in the central register.   & 0: No. 1: Yes\\
   \vspace{0.1cm} \\
            Q2 & Have you considered formally changing your address to increase the likelihood of getting admitted at your first prioritized high school?  & 0: No. 1: Yes \\
   \vspace{0.1cm} \\
            Q3 & Independently of whether you moved or not to increase the likelihood of getting admitted at your most preferred high school. What spoke against doing so (you can choose multiple answers) & 0: Nothing. 1: I was sufficiently certain that I would get accepted at my first priority. 2: I was afraid of getting caught by the authorities. 3: I had to contact my current school and tell them that I had moved. 4: I was worried of what my friends and family would think about me. 5: I knew no one I could move to with an address sufficiently close to my first priority high school. 6: I could not find a place to live sufficiently close to my first priority high school that I could afford. 7: I was worried what my prospective classmates would think about me. 8: Don't know. \\
			\hline\hline
		\end{tabularx}
\end{table}

\newpage
\begin{table}[ht!]\centering
	\def\sym#1{\ifmmode^{#1}\else\(^{#1}\)\fi}
	\caption*{\textbf{Table A.1 (continued):} Wording of survey questions}
		\begin{tabularx}{\textwidth}{l  X X }
			\hline\hline
			 Number&\multicolumn{1}{c}{Question} & \multicolumn{1}{c}{Scale} \\
			\hline
            Q4 & Independently of whether you moved or not to increase the likelihood of getting admitted at your most preferred high school. What spoke in favor doing so (you can choose multiple answers) & 0: My friends are enrolled at my first prioritized high school or guaranteed a seat. 1: My first prioritized high school is academically strong. 2: My first prioritized high school supplies exactly the courses, I want to take. 3: I believe, that I would fit in better at my first prioritized high school. 4: My first prioritized high school has nice social events. 5: I get a shorter commuting to my first prioritized high school. 6: I wanted to live in a more attractive neighborhood. 6: None of the abovementioned. 7: Don’t know. \\
	   \vspace{0.1cm} \\              		
            Q5 & Imagine that you have prioritized two high schools in your application. Also, imagine that you do not get admitted at your first prioritized high school. Do you believe that there is a higher, lower or unchanged likelihood of getting admitted at your second priority high school now than if you had ranked your second priority first?    & 0: Higher likelihood. 1: Unchanged likelihood. 2: Lower likelihood.  \\
   \vspace{0.1cm} \\
            Q6 &Do you know what determines who get admitted at which high school, if there are more applicants than vacant seats? (you can choose multiple answers) & 0: Attachment to the high school through siblings, who are either attending or have attended the school. 1: Distance to the high school. 2: Travel duration to the high school. 3: Match with the profile of the high school. 4: Grades. 5: Don’t know.   \\
 			\hline\hline
		\end{tabularx}
\end{table}

    \end{appendix}

\end{document}